\documentclass[a4paper,11pt]{article}
\usepackage[latin1]{inputenc}
\usepackage{amsmath,amsthm,amsfonts,amssymb}
\usepackage{enumerate}
\usepackage{graphicx}
\usepackage{float}

\usepackage{etex}
\usepackage{tikz}
\usetikzlibrary{decorations.markings}
\tikzset
{
   middlearrow/.style=
   {
   decoration={markings,mark= at position 0.5 with {\arrow{#1}} ,},
   postaction={decorate}
   }
}
\usepackage{nccmath}
\usepackage{mathtools}

\usepackage{authblk}

\usepackage[top=1.3in, bottom=1.3in, left=1.35in, right=1.35in]{geometry}
\usepackage{relsize}

\usepackage{array}

\theoremstyle{plain}
\newtheorem{theorem}{Theorem}[section]

\theoremstyle{definition}

\numberwithin{equation}{section}

\title{Analysis of a model for hepatitis C virus
 transmission that \\ includes the effects of vaccination with waning immunity}
\author[1,2$\ast$]{Daniah Tahir}
\author[2,3]{Abid Ali Lashari}
\author[4]{Kazeem Oare Okosun}
\affil[1]{Uppsala University, 75106, Uppsala, Sweden}
\affil[2]{National University of Sciences and Technology, H-12, Islamabad, Pakistan}
\affil[3]{Stockholm University, 10691, Stockholm, Sweden}
\affil[4]{Vaal University of Technology, Private Bag X021, Vanderbijlpark, South Africa}
\affil[$\ast$]{Correspondence to be sent to: Department of Mathematics, Uppsala University, 75106, Uppsala, Sweden. 
E-mail:~daniah.tahir@math.uu.se}
\date{}                    
\begin{document}

\maketitle

\begin{abstract} 
This paper considers a mathematical model based  on the transmission
dynamics of hepatitis C virus (HCV) infection. In addition to the
usual compartments for susceptible, exposed, and infected
individuals, this model includes compartments for individuals who
are under treatment and those who have had vaccination against HCV infection. It
is assumed that the immunity provided by the vaccine fades with
time. The  basic reproduction number, $R_0$, and the equilibrium
solutions of the model are determined.  The model exhibits the
phenomenon of backward bifurcation where a stable disease-free
equilibrium co-exists with a stable endemic equilibrium whenever $R_0$
is less than unity. It is shown that the use of only a perfect
vaccine can eliminate backward bifurcation completely. Furthermore, a
unique endemic equilibrium of the model is proved to be globally
asymptotically stable under certain restrictions on the parameter
values. Numerical simulation results are given to support the
theoretical predictions. 
[epidemiological model; equilibrium solutions; backward bifurcation; global asymptotic stability; Lyapunov function.]
\end{abstract}

\section{Introduction}

The liver of a hepatitis patient is one of the most frequently damaged organs in the body,
and it is indeed fortunate that it has a very large functional
reserve. In the experimental animal, it has been shown that only 10\% of the
 hepatic parenchyma (the functional part of the liver) is
required to maintain normal liver function (Cotran et al. 1994). The liver
can be infected due to a variety of infectious agents such as
parasites, viruses, and bacteria, and diseases of the liver have a
variety of causes such as obstructive, vascular, metabolic and toxic
involvements.

\noindent Hepatitis C is the inflammation of the liver caused by hepatitis C
virus (HCV), and spreads through contact with contaminated blood.
Hepatitis C may be an acute infection, which spans over a period of
weeks to a few months, or chronic infection, in which the virus
persists for a longer time (Di Bisceglie 2000; Das et al. 2005). Acute hepatitis is
characterized by moderate liver injury and if symptoms appear, they
include fatigue, loss of appetite, abdominal pain, fever and
jaundice. However, most of the times, acute hepatitis is
asymptomatic. A large percentage of patients with HCV infection
recover completely, but some develop long term chronic hepatitis
or massive necrosis of the liver. Chronic HCV infection may damage
the liver permanently, cause cirrhosis, hepatic failure, and
sometimes liver cancer.

\noindent Today, HCV infects an estimated 170 million people worldwide (Qesmi et al. 2010).
 Around 150 million people are chronically infected with
HCV. HCV infection is a major cause of death of more than 350,000
people every year. Countries with the highest prevalence of chronic
liver infection are Egypt (15\%), Pakistan (4.8\%) and China
(3.2\%) (Lozano 1990). Although, treatment for this infection does
exist, the current drug therapies are ineffective in completely
eliminating the virus and patients suffering from chronic illness
may require a liver transplant (Qesmi et al. 2010). Unfortunately, there is no
effective vaccine yet developed that may help prevent the spread of the disease. 
At present, various attempts are being made to create
such a vaccine (Chen and Li 2006). Thus, it is crucial to assess the
potential impact of HCV vaccine on the population.

\noindent Some mathematical models on HCV infection have been formulated
recently, but much work has not been done, since it is a relatively
new disease (discovered in 1989) and data is not available on
account of the high variability of the HCV. In contrast, more
research has been carried out on hepatitis B virus (HBV) infection.
Several epidemiological models have focused on the effects of
preventive measures as well as control of HBV infection (Zhang and Zhou 2012).
This has helped in creating cost effective disease prevention
techniques. The modes of transmission of both HCV and HBV are same,
i.e. through blood, thus mathematical models on both infections are
somewhat inter related. Some mathematical models were formed on HCV
infection that considered infected cells, uninfected cells and viral
cells in the human host. The basic aim of these models was to study
the effects of liver transplant in patients with HCV infection. But
in major cases, HCV infection is not completely eliminated even
after the transplant. Thus, these models were extended to include more infected
compartments (Dahari et al. 2005). Martcheva and Castillo-Chavez (2003) 
introduced an epidemiologic model of HCV infection
with chronic infectious stage in a varying population. Their model
does not include a recovered or immune class and it falls within the
susceptible-infected- susceptible (SIS) category of models. A
susceptible-infected-recovered (SIR) model was used by Jager et al. (2004) to study the transmission of HCV
among injecting drug users, while susceptible-infected-removed-susceptible
(SIRS) type models that allow waning immunity are presented in Zeiler et
al. (2010). Also, a deterministic model for HCV transmission is
used by Elbasha (2013), with the objective of assessing the impact of
therapy on public health.

\noindent Our aim is to meticulously analyze the model and examine various
parameters to explore their effect on the transmission of HCV and its
control. The model focuses on studying the effects of imperfect
vaccines on the control of hepatitis C. The model shows that an
imperfect vaccine reduces the number of individuals who are exposed
to HCV, while a perfect vaccine completely removes them. We have
subdivided the total population into six mutually-exclusive
compartments of susceptible, exposed, acutely infected, chronically
infected, treated and vaccinated individuals. Ordinary differential
equations are used to model the HCV infection.  This model can help 
provide insights into the spread of HCV infection and the assessment of 
the effectiveness of immunization techniques.

\noindent This paper is organized as follows: The mathematical model is
developed and analyzed in Section 2. The
stability of the disease free equilibrium, and the endemic equilibrium is discussed,
along with the effects of vaccination on backward bifurcation
phenomenon. Numerical simulations are also provided in the same section.
Section 3 summarizes the final results of the paper.

\section{Model Formulation}
\label{S2}

The total population at time $t$, denoted by $N(t)$, is divided into
sub-populations of susceptible individuals, $S(t)$, exposed
individuals with hepatitis C symptoms, $E(t)$, individuals with
acute infection, $I(t)$, individuals undergoing treatment, $T(t)$,
individuals with chronic infection, $C_h(t)$, and vaccinated
individuals, $V(t)$, so that
\[  
N(t)=S(t)+E(t)+I(t)+T(t)+C_h(t)+V(t).
\]            
It is assumed that the mode of transmission of HCV infection is
horizontal. We further assume that mixing of individual hosts is
homogeneous (every person in the population $N(t)$ has an equal
chance of getting HCV infection). The following system of ordinary
differential equations describes the dynamics of the HCV infection
\begin{equation}\label{eq2.1}
\begin{array}{rcl}
\displaystyle\frac{dS}{dt}&=&(1-b)\Lambda+\rho T+\alpha V-(\beta_{1}I+\beta _{2}C_{h}+\beta_{3}T)S +\sigma C_{h}-\mu S,\cr\cr
\displaystyle\frac{dE}{dt}&=&(\beta_{1}I+\beta_{2}C_{h}+\beta_{3}T)S+
(1-\psi)(\beta_{1}I+\beta_{2}C_{h}+\beta_{3}T)V-(\epsilon+\mu)E,\cr\cr
\displaystyle\frac{dI}{dt}&=&\epsilon E-(\kappa+\mu)I,\cr\cr
\displaystyle\frac{dT}{dt}&=&\pi_{1}\kappa I+\pi_{2}C_{h}-(\rho+\mu)T,\cr\cr
\displaystyle\frac{dC_{h}}{dt}&=&(1-\pi_{1})\kappa I-(\pi_{2}+\sigma+\mu)C_{h},\cr\cr 
\displaystyle\frac{dV}{dt}&=&b\Lambda-(\alpha+\mu)V-(1-\psi)(\beta _{1}I+\beta_{2}C_{h}+\beta_{3}T)V.
\end{array}
\end{equation}
The recruitment rate of susceptible humans is $\Lambda$. A
proportion, $b$, of these susceptible individuals is vaccinated. The
death rate of individuals is denoted by $\mu$. The rate of progression from acute
infected class to both treated and chronic infected class is given
by $\kappa$. The acutely infected proportion of individuals who
enter the treated class is $\pi_1$. The remaining infected
proportion, $(1-\pi_1)$, progresses to the chronic infectious stage. The
rate of progression for treatment from chronic hepatitis is given by
$\pi_2$. The term $ \epsilon $ is the rate of progression from
exposed class to acute infected class. The recovery rates due to
treatment and naturally from the chronic group are $\rho$ and
$\sigma$, respectively.

\noindent The transmission coefficients of HCV infection by individuals
with acute hepatitis C, $I(t)$, chronic hepatitis C, $C_h(t)$ and
individuals undergoing treatment but not yet cured, $T(t)$ are
$\beta_1, \beta_2,$ and $\beta_3$, respectively. Following effective
contact with $I(t)$, $C_(t)$, and $T(t)$, susceptible individuals
can acquire HCV at a rate $(\beta_1 I + \beta_2 C_h + \beta_3 T)$.
$\psi$ ($0<\psi\leq1$) represents the vaccine efficacy, with $\psi=1$
representing a perfect vaccine, and $\psi \in (0,1) $ corresponding to
an imperfect vaccine which wanes with time. The term $(1-\psi)$
corresponds to the decrease in disease transmission in vaccinated
individuals, in contrast to susceptible individuals who are not
vaccinated. Hence, vaccinated individuals acquire HCV at a reduced
rate $(1-\psi)(\beta_1 I + \beta_2 C_h + \beta_3 T)$. The rate at
which the vaccine wanes is denoted by $\alpha$. The parameter
description is described in Table \ref{tab1}.

\begin{table}
\centering
\caption{Description of parameters} 
\label{tab1}
\begin{tabular}{lll}
\hline\noalign{\smallskip}
Parameter & Description  \\
\noalign{\smallskip}\hline\noalign{\smallskip}
$\Lambda$ & recruitment rate of individuals  \\
$\mu$           & death rate of individuals    \\
$\alpha$ & waning rate of the vaccine  \\
$\psi$ & vaccine efficacy \\
$\beta_i$           & transmission rate (i=1, 2, 3)  \\
$b$     & proportion of vaccinated individuals   \\
$\kappa$  & rate of progression from the acute state &\\& to treated and
chronic state
\\
$\epsilon$  & rate of transfer from exposed class &\\&
 to acute
infected class \\
$\pi_1$ &  proportion of individuals who enter &\\& 
the treated class from acutely infected class \\
 $\pi_2$ &  rate of progression
for treatment&\\& from chronic hepatitis\\
$\rho$ & rate of recovery due to treatment \\
$\sigma$ & rate of recovery from the chronic class \hspace{0.5cm}\\
\noalign{\smallskip}\hline
\end{tabular}
\end{table}

\noindent In the proposed model (\ref{eq2.1}), the total population is $ S +
E + I + T + C_h + V = \frac{\Lambda}{\mu}\ $ for all $t\geq0$,
provided that $S(0) + E(0) + I(0) + T(0) + C_h(0) + V(0) = \frac{
\Lambda}{\mu}.$ Thus, the biologically feasible region for system
(\ref{eq2.1}) given by
\[
\Delta= \{(S,E,I,T,C_h,V)\in
R^6:S+E+I+T+C_h+V=\displaystyle \frac{\Lambda}{\mu}\},
\]
is positively invariant with respect to the system (\ref{eq2.1}).


\subsection{Local stability of the disease-free equilibrium (DFE)}
\label{S3} 

For the mathematical model in equation (\ref{eq2.1}), the DFE, $P_0$ is given by
\[
(S_0,E_0,I_0,T_0,C_{h0},V_0)= 
\Big(\displaystyle
\frac{(1-b)\Lambda}{\mu}+\frac{\alpha
b\Lambda}{\mu(\alpha+\mu)},0,0,0,0,\displaystyle\frac{b\Lambda}{\alpha+\mu}\Big).
\]
The local stability of $P_0$ is determined by the next
generation operator method (Driessche and Watmough 2002) on system (\ref{eq2.1}).
For this purpose, the basic reproduction number (the average number
of secondary infections produced by an infected individual in a
completely susceptible population), denoted by  $R_0$, is obtained.
Using the same notation as in Driessche and Watmough (2002), 
$R_0$ 
is given by 
\[
R_0= \frac
{\epsilon}{K_1K_2 }(\displaystyle
\frac{(1-b)\Lambda}{\mu}+\frac{\alpha b\Lambda}{\mu
K_5}+(1-\psi)\displaystyle\frac{b\Lambda}{K_5})\Bigr[ \beta_1 +
\beta_2 \frac{\kappa(1 - \pi_1)}{ K_4 } + \beta_3 \frac{\left( \pi_1
\kappa K_4 + \pi_2 \kappa(1 - \pi_1)\right)}{K_3K_4}\Bigr],
\]
where
\[
K_1= \epsilon+\mu,~ K_2=\kappa+\mu,~ K_3=\rho+\mu,
~K_4=\pi_2+\sigma+\mu, K_5=\alpha+\mu.
\]
Using Theorem 2 in Driessche and Watmough (2002), the
following result is established.

\begin{theorem}
The DFE of the model (\ref{eq2.1}), is locally asymptotically stable if $R_0 < 1$, and unstable if $R_0 > 1$.
\end{theorem}

\subsection{Endemic equilibria and backward bifurcation}
\label{S4}

To calculate the endemic equilibrium, we consider the following reduced system of
differential equations
\begin{equation}
\label{eq3.2}
\begin{array}{rcl}
\displaystyle\frac{dE}{dt}&=&(\beta_{1}I+\beta_{2}C_{h}+\beta_{3}T)(\displaystyle\frac{\Lambda}{\mu}
-E-I-T-C_h-\psi V)-K_1E,\cr\cr \displaystyle\frac{dI}{dt} &=&
\epsilon E-K_2I,\cr\cr \displaystyle\frac{dT}{dt}&=&\pi_{1}\kappa
I+\pi_{2}C_{h}-K_3T,\cr\cr
\displaystyle\frac{dC_h}{dt}&=&(1-\pi_{1})\kappa I-K_4C_{h},\cr\cr
\displaystyle\frac{dV}{dt} &=& b\Lambda-K_5V-(1-\psi)(\beta
_{1}I+\beta _{2}C_{h}+\beta_{3}T)V .
\end{array}
\end{equation}
We will consider the dynamics of the flow generated by (\ref{eq3.2})
in the invariant region 
\[
\Omega= \{E +I +T +C_h + V \leq
\displaystyle\frac{\Lambda}{\mu}\}.
\]
The endemic equilibrium for system (\ref{eq3.2}) is
$P^{*}(E^{\ast},I^{\ast},T^{\ast },C_{h}^{\ast },V^{\ast })$, where

\begin{equation}
\label{eq3.3}
\begin{array}{rcl}
E^{\ast }&=&\displaystyle\frac{K_2 I^\ast }{\epsilon},\cr\cr T^{\ast
}&=&\displaystyle \frac{(\pi_1 \kappa K_4 +
\pi_2(1-\pi_1)\kappa)I^\ast }{K_4K_3},\cr\cr C_{h}^{\ast
}&=&\displaystyle\frac{(1-\pi_1)\kappa I^\ast }{K_4},\cr\cr
V^\ast&=&\displaystyle\frac{b\Lambda}{K_5 + (1-\psi) [ \beta_1 +
\beta_2 \displaystyle\frac{\kappa (1 - \pi_1)}{ K_4 } + \beta_3
\frac{\left( \pi_1 \kappa K_4 + \pi_2 \kappa(1 -
\pi_1)\right)}{K_3K_4}] I^\ast},
\end{array}
\end{equation}
and $I^\ast$ is the root of the following quadratic equation

\begin{equation}
\label{eq3.4}
a_1 I^{\ast 2}+a_2I^{\ast}+a_3=0.
\end{equation} 

\noindent Here

\begin{equation}
\label{eq3.4a}
\begin{array}{rcl}
a_1&=&(1-\psi) B^2[\mu K_2K_3K_4+\epsilon\mu\ K_3K_4
+\left(\pi_1 \kappa K_4+\pi_2
\kappa(1-\pi_1)\right)\epsilon\mu\cr\cr&& +\epsilon
\kappa\mu(1-\pi_1)K_3],\cr\cr 
a_2&=&B[\mu K_2K_3K_4K_5+\epsilon\mu
K_3K_4K_5+ \epsilon\mu K_5(\pi_1 \kappa K_4+
\pi_2(1-\pi_1)\kappa)\cr\cr&& + (1-\pi_1)\epsilon\kappa\mu K_3K_5
+(1-\psi)\mu K_1K_2K_3K_4  -(1-\psi)\Lambda\epsilon BK_3K_4],\cr\cr
a_3&=&\mu K_1K_2K_4K_3K_5(1-R_0),
\end{array}
\end{equation}
with
\[
 B=\bigg[\beta_1 + \beta_2 \frac{\kappa (1 - \pi_1)}{ K_4 } + \beta_3 \frac{( \pi_1 \kappa K_4 + \pi_2 \kappa(1 -
\pi_1))}{K_3K_4}\bigg].
\]
The endemic equilibria of the model (\ref{eq3.2}) can then be
obtained by solving for $I^\ast$ from (\ref{eq3.4}), and
substituting the positive values of $I^\ast$ into the expressions in
(\ref{eq3.3}). Hence, $S^\ast $ can be determined from
$\frac{\Lambda}{\mu}-E^\ast-I^\ast-T^\ast-C_h^\ast-V^\ast$. From
(\ref{eq3.4a}), it can be seen that $a_1$ is always positive (for an
imperfect vaccine), and $a_3$ is positive (negative) if $R_0$ is
less than (greater than) unity. Thus, the following
result is established
\begin{theorem} The model in (\ref{eq3.2}) has:

(i) a unique endemic equilibrium if  $a_3 < 0 \Leftrightarrow R_0>
1$;

(ii) a unique endemic equilibrium if $a_2 < 0$, and $a_3=0$ or
$a_2^2 - 4a_1a_3 = 0$;

(iii) two endemic equilibria if $a_3 > 0, a_2 < 0 $ and  $a_2^2 -
4a_1a_3 > 0$;

(iv) no endemic equilibrium otherwise.
\end{theorem}

\begin{figure*}[!t]
\centerline{
\includegraphics[width=.7\textwidth]{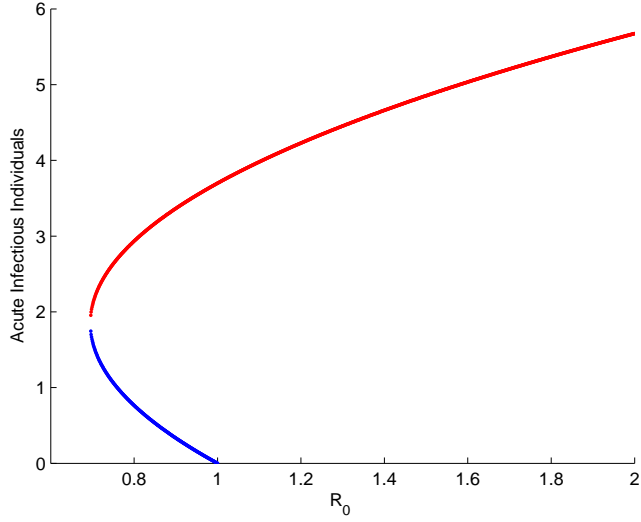}}
\caption{Backward Bifurcation diagram, with parameter values
$\beta_2=0.09$, $\beta_3=0.19$, $\mu=0.00004$, $\alpha=0.1$,
$\rho=0.152$, $\pi_1=0.001$, $\pi_2=0.02$, $\epsilon=0.022$,
$\kappa=0.032$, $\Lambda=0.0052$, $\sigma=0.2$,
$\psi=0.95$.}
\label{bbn}
\end{figure*}

\noindent Hence, the model has a unique endemic equilibrium ($P^*$) whenever
$R_0>1$, as evident from case (i) of the above theorem. Also, case
(iii)  indicates a possible chance of backward bifurcation (where a locally asymptotically stable DFE exists along with a
locally asymptotically stable endemic equilibrium when $R_0<1$).
Since, for $a_3>0$, $R_0<1$, the model will have a disease-free
equilibrium and two endemic equilibria.  To check for this, the
discriminant $a_2^2 - 4a_1a_3$ is set to zero and is solved for the
critical value of $R_0$. The critical value is denoted by $R_c$ and is given by
\[
R_c= 1 - \frac{a_2^2}{4a_1 \mu K_1K_2K_5K_3K_4}.
\]
Backward bifurcation occurs for those values of $R_0$ which satisfy $R_c
< R_0 < 1$. This is illustrated by simulating the model with these
parameter values: $\beta_1=0.03$, $\beta_3=0.19$, $\mu=0.00004$,
$\alpha=0.1$, $\rho=0.152$, $\pi_1=0.001$, $\pi_2=0.02$,
$\epsilon=0.022$, $\kappa=0.032$, $\Lambda=0.0052$, $\sigma=0.2$,
$\psi=0.95$. (These values are used merely for illustration
purposes, and may not be realistic from epidemiological point of
view.) The result is shown in Fig. \ref{bbn}. It can be seen that a
locally asymptotically stable disease free equilibrium, a locally
asymptotically stable endemic equilibrium, and, an unstable endemic
equilibrium coexist when $R_0 < 1$.

\subsubsection{Proof of backward bifurcation phenomenon}
\label{S5} 

The phenomenon of backward bifurcation can be proved by using the center
manifold theory on system (\ref{eq2.1}). A theorem given by Castillo-Chavez
 and Song (2004) will be used here. To apply this
method, the following change of variables is made on the model:
\[
x_1=S, x_2=E, x_3=I, x_4=T, x_5=C_h, x_6= V.
\]
Let
\[
X=(x_1,x_2,x_3,x_4,x_5,x_6)^T.
\]
Thus, the system (\ref{eq2.1}) can now
be written as $\frac{dX}{dt}=(f_1,f_2,f_3,f_4,f_5,f_5,f_6)^T$ and is given
below

\begin{equation}
\label{eq3.7}
\begin{array}{rcl}
\displaystyle\frac{dx_1}{dt}&=f_1=&(1-b)\Lambda+\rho x_4+\alpha
x_6-(\beta_{1}x_3+\beta _{2}x_5+\beta_{3}x_4)x_1+\sigma x_5-\mu x_1,\cr\cr
\displaystyle\frac{dx_2}{dt}&=f_2=&(\beta_{1}x_3+\beta_{2}x_5+\beta_{3}x_4)x_1+(1-\psi)(\beta_{1}x_3+\beta_{2}x_5+\beta_{3}x_4)
x_6-K_1x_2,\cr\cr \displaystyle\frac{dx_3}{dt} &=f_3=& \epsilon
x_2-K_2x_3,\cr\cr
\displaystyle\frac{dx_4}{dt}&=f_4=&\pi_{1}\kappa x_3+\pi_{2}x_5-K_3x_4,\cr\cr
\displaystyle\frac{dx_5}{dt}&=f_5=&(1-\pi_{1})\kappa x_3-K_4x_5,\cr\cr
\displaystyle\frac{dx_6}{dt} &=f_6=& b\Lambda-K_5x_6-(1-\psi)(\beta
_{1}x_3+\beta _{2}x_5+\beta_{3}x_4)x_6.
\end{array}
\end{equation}
Choose $\beta_1$ as the bifurcation parameter, and let $R_0=1$.
Solving for  $\beta_1=\bar\beta_1$ from $R_0=1$ gives
\[ 
\beta_1=\bar\beta_1=\\~\\\displaystyle\frac{K_1K_2}{\epsilon
A}-\frac{\beta_2(1-\pi_1)\kappa}{K_4}-\frac{\beta_3(\pi_1\kappa K_4+\pi_2(1-\pi_1)\kappa)}{K_3K_4}
\]
where
\[
 A = \displaystyle
\frac{(1-b)\Lambda}{\mu}+\frac{\alpha b\Lambda}{\mu
K_5}+(1-\psi)\displaystyle\frac{b\Lambda}{K_5}.
\]
The Jacobian matrix ($\textbf{J}$) of system
(\ref{eq3.7}) calculated at $P_0$, with $\beta_1=\bar
\beta_1$, is given as follows
\[
\textbf{J}\\=\left(\begin{array}{cccccc} -\mu\vspace{.5mm} & 0\vspace{.5mm} &
-\beta_1K_h\vspace{.5mm}
&\rho-\beta_3K_h\vspace{.5mm}&\sigma-\beta_2
K_h\vspace{.5mm}&\alpha
\vspace{.5mm}\cr 0\vspace{.5mm} & -K_1\vspace{.5mm} & \beta_1 A
\vspace{.5mm}& \beta_3 A \vspace{.5mm}&\beta_2 A\vspace{.5mm}&0 \cr
0\vspace{.5mm} & \epsilon \vspace{.5mm}& -K_2 \vspace{.5mm}&
0\vspace{.5mm}&0\vspace{.5mm}&0 \cr 0 \vspace{.5mm}& 0
\vspace{.5mm}& \pi_1 \kappa \vspace{.5mm}& -K_3 \vspace{.5mm}& \pi_2
\vspace{.5mm}& 0\cr
0\vspace{.5mm}&0\vspace{.5mm}&(1-\pi_1)\kappa\vspace{.5mm}&0\vspace{.5mm}&-K_4\vspace{.5mm}&0\cr
0&0&-\beta_1K_m&-\beta_3K_m&-\beta_2K_m&-K_5
\end{array}\right),
\]
where
\[K_h=(\displaystyle\frac{\Lambda}{\mu}-\displaystyle\frac{b\Lambda}{K_5}),  K_m=\displaystyle\frac{(1-\psi)\Lambda
b}{K_5}.
\]
The characteristic equation (in $\lambda$) of $\textbf{J}$ is given as

\begin{equation}
\label{j}
(-\mu-\lambda)(-K_5-\lambda)\big(\lambda^4+D_1\lambda^3
+D_2\lambda^2+D_3\lambda+D_4 \big)=0,
\end{equation}
where
\[
\begin{array}{rcl}
D_1&=&K_1+K_2+K_3+K_4,\cr\cr
D_2&=&K_3K_4+K_1K_3+K_2K_3+K_1K_4+K_2K_4+K_1K_2-\beta_1\epsilon
A,\cr\cr
D_3&=&K_1K_3K_4+K_2K_3K_4+K_1K_2K_3+K_1K_2K_4-\beta_1\epsilon
A(K_3+K_4)\cr\cr &&-\beta_3 \kappa\pi_1 \epsilon A-
(1-\pi_1)\beta_2\kappa\epsilon A, \cr\cr D_4&=&K_1K_2K_3K_4(1-R_0).
\end{array}
\]
For $R_0=1$, the characteristic equation (\ref{j}) becomes

\begin{equation}\label{ch}
\lambda(-\mu-\lambda)(-K_5-\lambda)\big(\lambda^3+D_1\lambda^2
+D_2\lambda+D_3\big)=0.
\end{equation}

\noindent Equation (\ref{ch}) has a zero eigenvalue and two
negative eigenvalues, $-\mu$ and $-K_5$. The remaining three
eigenvalues are given by the following cubic equation in $\lambda$

\begin{equation}\label{cubic}
\lambda^3+D_1\lambda^2 +D_2\lambda+D_3=0.
\end{equation}

\noindent $D_1$ is clearly positive. $D_2$ and
$D_3$ can easily be shown to be positive when $\beta_1$ is replaced
with $\bar \beta_1$. Similarly, $D_1D_2-D_3>0$. Hence, using the
Routh-Hurwitz criterion (Allen 2007), all roots of the
characteristic equation (\ref{cubic}) have negative real parts.
Therefore, the Jacobian matrix of the linearized system has a simple
zero eigenvalue, with all other eigenvalues having negative real
parts. Hence, the Center Manifold Theory (Castillo-Chavez and Song 2004) can be used
to analyze the dynamics of system (\ref{eq3.7}).

\noindent Corresponding to the zero eigenvalue, the Jacobian matrix
$\textbf{J}\mid_{\beta_1=\bar\beta_1}$ can be shown to have a right
eigenvector given by $w=(w_1,w_2,w_3,w_4,w_5,w_6)^T$,
where
\[
\begin{array}{rcl}
w_1&=&\displaystyle\frac{1}{\mu}\Bigr(\rho(\displaystyle\frac{\pi_1\kappa
K_4+\pi_2(1-\pi_1)\kappa}{K_3K_4})+\sigma\displaystyle\frac{(1-\pi_1)\kappa}{K_4}-
\displaystyle\frac{K_1K_2}{\epsilon A}(\displaystyle
\frac{(1-b)\Lambda}{\mu}+\frac{\alpha b\Lambda}{\mu
K_5} \cr\cr &&+\displaystyle\frac{\alpha(1-\psi) b\Lambda}{K_5^2})\Bigr)w_3,\cr\cr
w_2&=&\displaystyle\frac{K_2}{\epsilon}w_3,\cr\cr
w_3&=&w_3,\cr\cr
w_4&=&\displaystyle\frac{\pi_1 \kappa}{K_3}w_3 +
\displaystyle\frac{\pi_2(1-\pi_1)\kappa}{K_3K_4}w_3,\cr\cr
w_5&=&\displaystyle\frac{(1-\pi_1)\kappa}{K_4}w_3,\cr\cr
w_6&=&\displaystyle\frac{-(1-\psi) b\Lambda K_1K_2}{K_5^2 \epsilon
A}w_3.
\end{array}
\]
Similarly, corresponding to the zero
eigenvalue, $\textbf{J}\mid_{\beta_1=\bar\beta_1}$ has a left eigenvector
given by $v=(v_1,v_2,v_3,v_4,v_5,v_6)$, where
\[
\begin{array}{rcl}
v_1&=&0,\cr\cr
v_2&=&\displaystyle\frac{\epsilon}{K_1}v_3,\cr\cr
v_3&=&v_3,\cr\cr
v_4&=&\displaystyle\frac{\epsilon\beta_3 A}{K_1K_3}v_3,\cr\cr
v_5&=&\displaystyle\frac{\epsilon\beta_2 A}{K_1K_4}v_3 +
\displaystyle\frac{\epsilon\beta_3\pi_2 A}{K_1K_4K_3}v_3,\cr\cr
v_6&=&0.
\end{array}
\]

\vspace{.2cm}

\noindent \textbf{Calculation of a}. For system (\ref{eq3.7}), the
corresponding non-zero partial derivatives of $f_i$ $(i=1,2,...,6)$
calculated at the DFE, $P_0$, are given by
\[
\displaystyle\frac{\partial^2 f_1}{\partial
 x_1\partial x_3}=-\beta_1, \displaystyle\frac{\partial^2 f_1}{\partial x_1
\partial x_5}=-\beta_2,
\displaystyle\frac{\partial^2 f_1}{\partial x_1
\partial x_4}=-\beta_3,
\displaystyle\frac{\partial^2 f_2}{\partial x_1\partial
x_3}=\beta_1, \displaystyle\frac{\partial^2
f_2}{\partial x_1\partial x_5}=\beta_2,
\]
\[
\displaystyle\frac{\partial^2 f_2}{\partial x_1\partial
x_4}=\beta_3, \displaystyle\frac{\partial^2
f_2}{\partial x_3\partial x_6}=(1-\psi)\beta_1,
\displaystyle\frac{\partial^2 f_2}{\partial x_5
\partial x_6}=(1-\psi)\beta_2, \displaystyle\frac{\partial^2 f_2}{\partial x_4\partial
x_6}=(1-\psi)\beta_3,
\]
\[
 \displaystyle\frac{\partial^2
f_6}{\partial x_3\partial x_6}=-(1-\psi)\beta_1,
\displaystyle\frac{\partial^2 f_6}{\partial x_3\partial
x_6}=-(1-\psi)\beta_2, \displaystyle\frac{\partial^2
f_6}{\partial x_4\partial x_6}=-(1-\psi)\beta_3.
\]

\vspace{.2cm}

\noindent Consequently, the
associated bifurcation coefficient, a, is given by
\[
\begin{array}{rcl} 
\textrm{a}&=&\displaystyle\sum_{k, i, j=1}^6 u_k w_i
w_j\displaystyle\frac{\partial^2f_k}{\partial y_i \partial y_j}
(0,0)\cr\cr  &=&\displaystyle\frac{\epsilon u_3w_3^2}{K_1 \mu}\left(\beta_1 +
\beta_2 \displaystyle\frac{\kappa (1 - \pi_1)}{ K_4 } + \beta_3
\displaystyle\frac{( \pi_1 \kappa K_4 + \pi_2 \kappa(1 -
\pi_1))}{K_3K_4}\right)\Bigr[\rho\left(\displaystyle\frac{\pi_1\kappa
K_4+\pi_2(1-\pi_1)\kappa}{K_3K_4}\right)\cr\cr
&&+\sigma\displaystyle\frac{(1-\pi_1)\kappa}{K_4}-
\displaystyle\frac{K_1K_2}{\epsilon A}\left(\displaystyle
\frac{(1-b)\Lambda}{\mu}+\frac{\alpha b\Lambda}{\mu
K_5}+\frac{\alpha(1-\psi)
b\Lambda}{K_5^2}\right)-\displaystyle\frac{(1-\psi)^2\mu b\Lambda
K_1K_2}{K_5^2 \epsilon A}\Bigr].
\end{array}
\]

\vspace{.2cm}

\noindent \textbf{Calculation of b}. The required partial derivative, for the
computation of b, is calculated at $P_0$, and is given by
 $\displaystyle\frac{\partial^2 f_2}{\partial
 x_3\partial \beta_1}= A $.
Hence, the associated bifurcation coefficient, b, is given as
\[
\begin{array}{rcl}
\textrm{ b}&=& \displaystyle\sum_{k, i=1}^6 u_k
w_i\displaystyle\frac{\partial^2f_k}{\partial x_i\partial \phi}
(0,0)=\displaystyle\frac{A\epsilon u_3 w_3}{K_1} > 0 .
\end{array}
\]
Since the coefficient b is always positive, it follows from
Theorem 3.3 given by Castillo-Chavez and Song B (2004) that the system (\ref{eq3.2}) will undergo backward
bifurcation if the coefficient a is positive.

\noindent The phenomenon of backward bifurcation poses a lot of
problems, since it jeopardizes the possibility of total disease
eradication from the population, when the basic reproduction number
is less than unity. Hence, it is instructive to try to eliminate the
backward bifurcation effect. Since, this effect requires the
existence of at least two endemic equilibria when $R_0 < 1$
(Garba et al. 2008; Safi and Gumel 2011), it may be removed by considering such a model in
which positive endemic equilibria cease to exist.

\subsubsection{Use of a perfect vaccine to eliminate backward bifurcation}
\label{S6}

The backward bifurcation behavior of the proposed HCV infection
model (\ref{eq2.1}), can be eliminated by using a perfect vaccine,
i.e., when $\psi$=1. For $\psi$=1, the original model now becomes
\begin{equation}\label{eq3.8}
\begin{array}{rcl}
\displaystyle\frac{dS}{dt} &=&(1-b)\Lambda+\rho T+\alpha
V-(\beta_{1}I+\beta _{2}C_{h}+\beta_{3}T)S+\sigma C_{h}-\mu S,\cr\cr
\displaystyle\frac{dE}{dt}&=&(\beta_{1}I+\beta_{2}C_{h}+\beta_{3}T)S-K_1E,\cr\cr
\displaystyle\frac{dI}{dt} &=& \epsilon E-K_2I,\cr\cr
\displaystyle\frac{dT}{dt}&=&\pi_{1}\kappa
I+\pi_{2}C_{h}-K_3T,\cr\cr
\displaystyle\frac{dC_h}{dt}&=&(1-\pi_{1})\kappa I-K_4C_{h},\cr\cr
\displaystyle\frac{dV}{dt} &=& b\Lambda-K_5V.
\end{array}
\end{equation}

\noindent  System (\ref{eq3.8}) has a DFE, $P_0
(S_0,0,0,0,0,V_0)$, which is the same as the original model given in
equation (\ref{eq2.1}). The corresponding \emph{vaccinated
reproduction number}, $\bar R_0 $, for model (\ref{eq3.8}) is given
as
\[\bar R_0 = R_0\mid_{ \psi=1} = \frac {\epsilon}{K_1 K_2 }\bigg(\displaystyle \frac{(1-b)\Lambda}{\mu}+\frac{\alpha b\Lambda}{\mu
K_5}\bigg)\bigg( \beta_1 + \beta_2 \frac{\kappa (1 - \pi_1)}{ K_4 }
+ \beta_3 \frac{\left( \pi_1 \kappa K_4 + \pi_2 \kappa(1 -
\pi_1)\right)}{K_3K_4}\bigg).
\]
Consider now, the quadratic equation (\ref{eq3.4}), rewritten below for
convenience
\[
a_1 I^{\ast 2}+a_2I^{\ast}+a_3=0.
\]
For $\psi=1$, using the values given in equation (\ref{eq3.4a}), the
coefficients $a_1, a_2$, and $a_3$ of the above quadratic equation
reduce to $a_1=0,$ $a_2>0,$ and $a_3\geq 0$ (whenever $\bar R_0=
R_0\mid_{ \psi=1}\leq1$). In this case, the quadratic equation
(\ref{eq3.4}) will have just a single non positive solution
\[
I^*=-\displaystyle\frac{a_3}{a_2}\leq0.
\]
Hence, whenever $\bar R_0 \leq1$, the model
(\ref{eq3.8}) has no positive endemic
equilibrium. This clearly suggests the impossibility of backward
bifurcation (because for backward bifurcation to occur, there must
exist at least two endemic equilibria whenever $\bar R_0 \leq1$).

\begin{table}
\caption{Values of parameters} 
\label{tab2}
\begin{tabular}{llll}
\hline\noalign{\smallskip}
Parameter  & Value(range) & Units & Source\\
\noalign{\smallskip}\hline\noalign{\smallskip}
$\Lambda$ & 85 &per year& (Martin et al. 2011; Martin et al. 2011)  \\
$\mu$           & 0.085 & per year&   (Martin et al. 2011; Martin et al. 2011) \\
$\beta_i$           & (0,1) & per year& (Martin et al. 2011; Martin et al. 2011)  \\
$\pi_1$ &  0.26 &- & (Martin et al. 2011; Martin et al. 2011)\\
$\rho$ & 1.992& per year &(Martin et al. 2011)\\
$\psi$ & (0,1] & -&Variable \\
$\alpha$ & 0.006 &- &Assumed \\
$b$     & 0.4&-& Assumed   \\
$\kappa$  & 2.085& -& Assumed
\\
$\epsilon$  & 0.569&-&Assumed\\
 $\pi_2$ & 0.25 & -&Assumed\\
$\sigma$ &  0.004 &-&Assumed \hspace{0.5cm}\\
\noalign{\smallskip}\hline
\end{tabular}
\end{table}

\begin{figure*}[!t]
\centerline{
\includegraphics[scale=0.7]{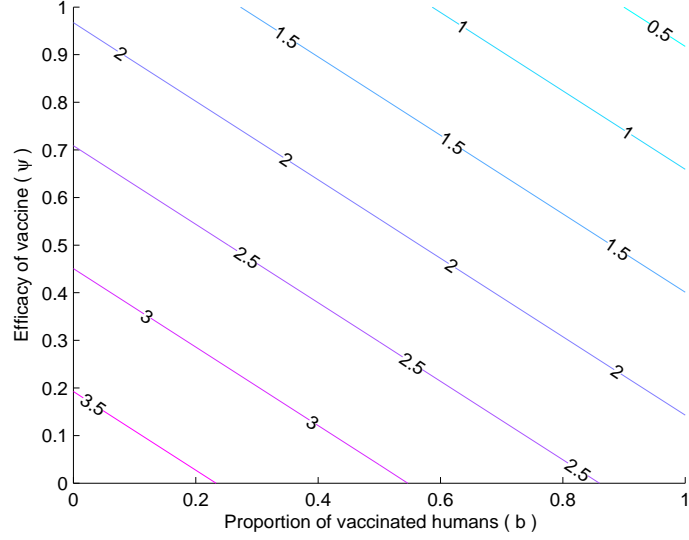}}
\caption{Simulation of the model (\ref{eq3.8}), showing a contour
plot of $\bar R_0$ as a function of proportion of vaccinated humans
($b$) and vaccine efficacy ($\psi$).}
\label{vacc}
\end{figure*}

\noindent A contour plot of vaccinated reproduction number ($\bar R_0$) as a
function of proportion of vaccinated humans ($b$) and vaccine
efficacy ($\psi$) is shown in Fig. \ref{vacc}. The parameter values
used to generate this diagram are given in Table \ref{tab2}. The
contours illustrate a significant decrease in the vaccinated
reproduction number, $\bar R_0$, with increasing vaccine efficacy,
$\psi$,  and proportion of vaccinated humans, $b$. It can be seen
that very high vaccine efficacy and vaccine coverage is required to
control HCV infection effectively in the population. Almost all of
the susceptible individuals should have had vaccination, and vaccine
efficacy must be 100\% for $\bar R_0$ to be less than one, so that
the spread of HCV infection is controlled effectively.

\noindent The global stability of the disease free
equilibrium can be proved in the region $\Delta$ as follows.

\begin{theorem} For a perfect vaccine ($\psi=1$),  $P_0$  is globally
asymptotically stable in $\Delta$ whenever
\[
\bar R_0\leq \displaystyle\frac{S_0 \mu}{\Lambda}<1,
\]
where
\[
S_0=\displaystyle \frac{(1-b)\Lambda}{\mu}+\frac{\alpha
b\Lambda}{\mu(\alpha+\mu)}.
\]
\end{theorem}

\begin{figure*}[!t]
\centering
\includegraphics[width=0.49\textwidth]{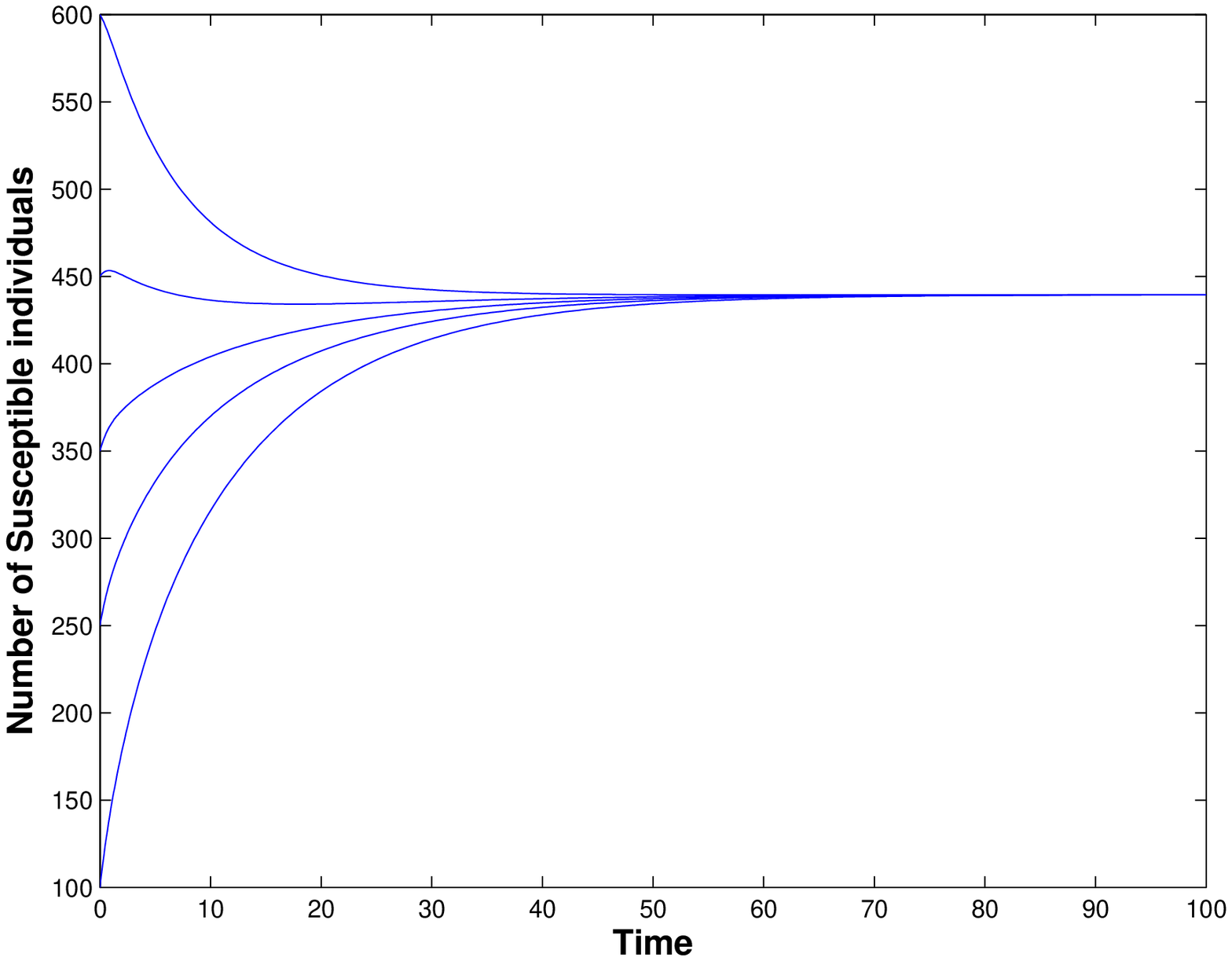}
\includegraphics[width=0.49\textwidth]{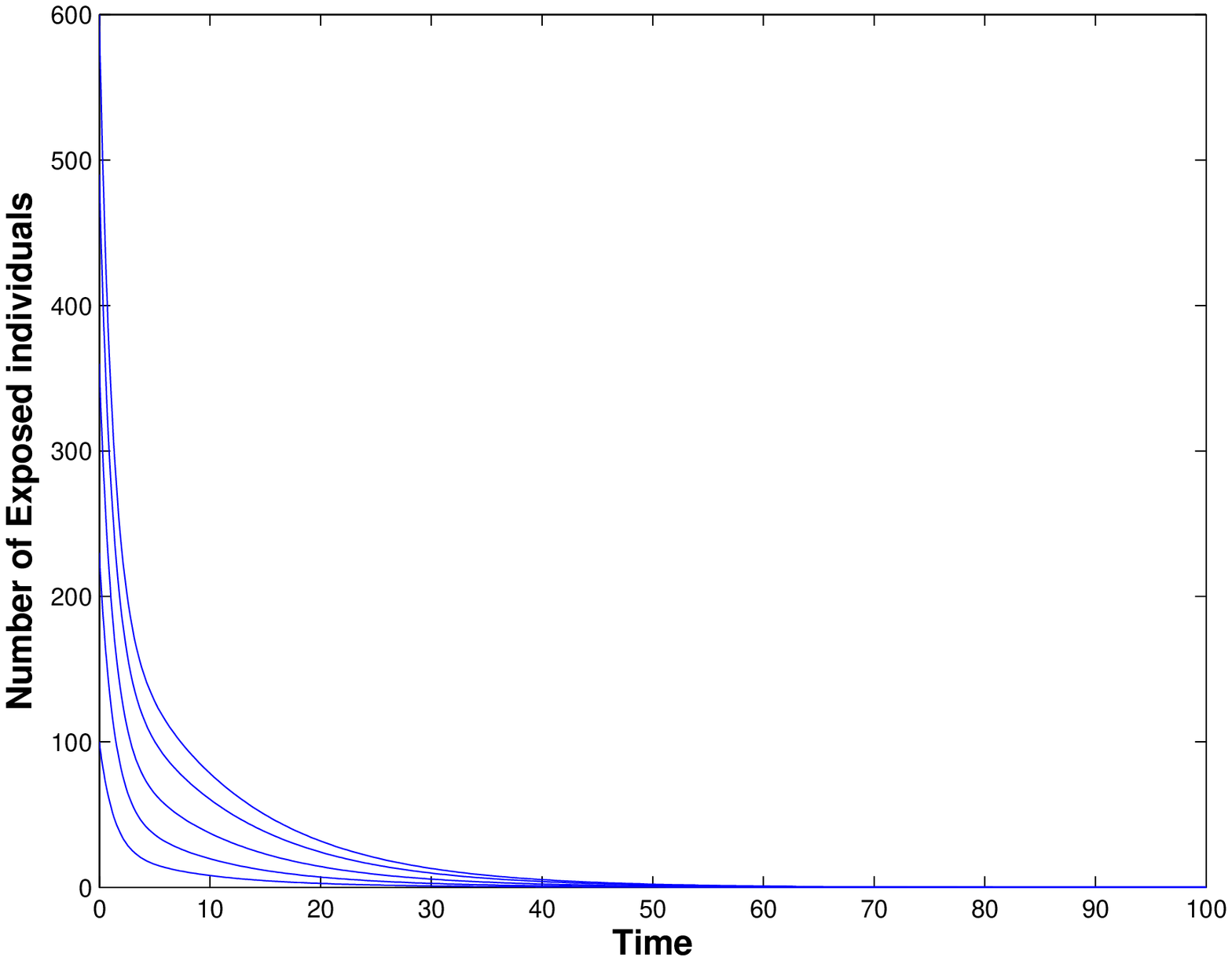}
\includegraphics[width=0.49\textwidth]{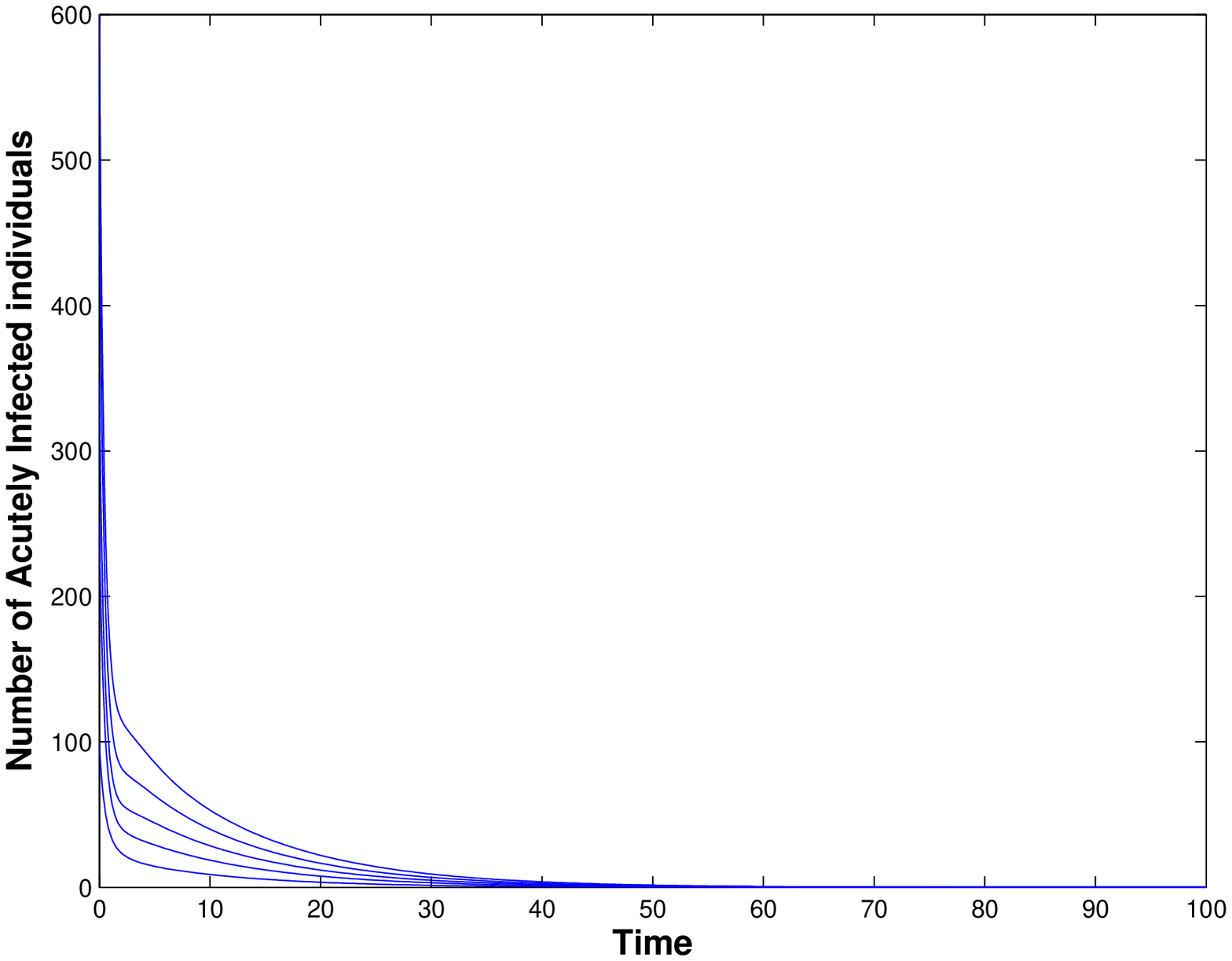}
\includegraphics[width=0.49\textwidth]{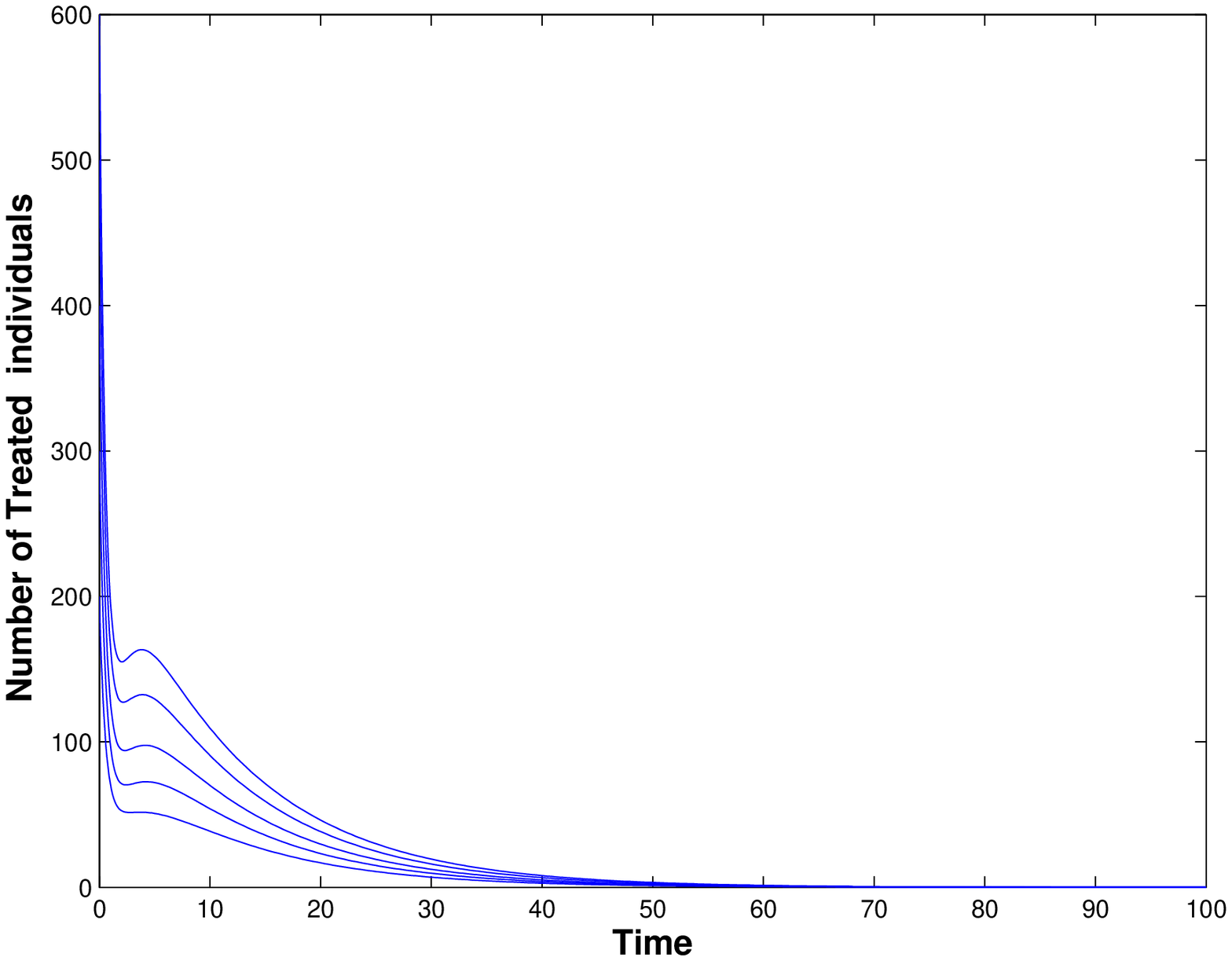}
\includegraphics[width=0.49\textwidth]{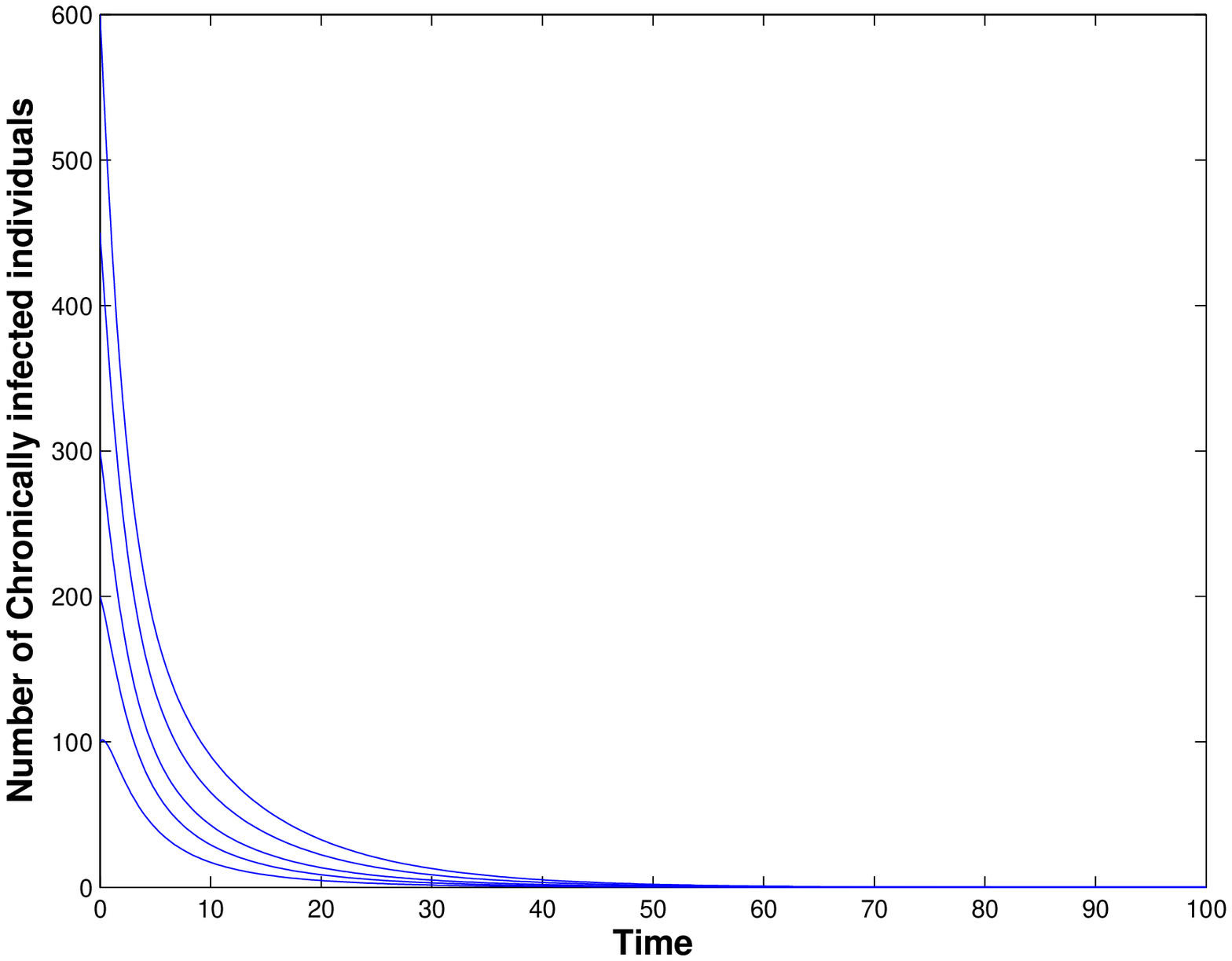}
\includegraphics[width=0.49\textwidth]{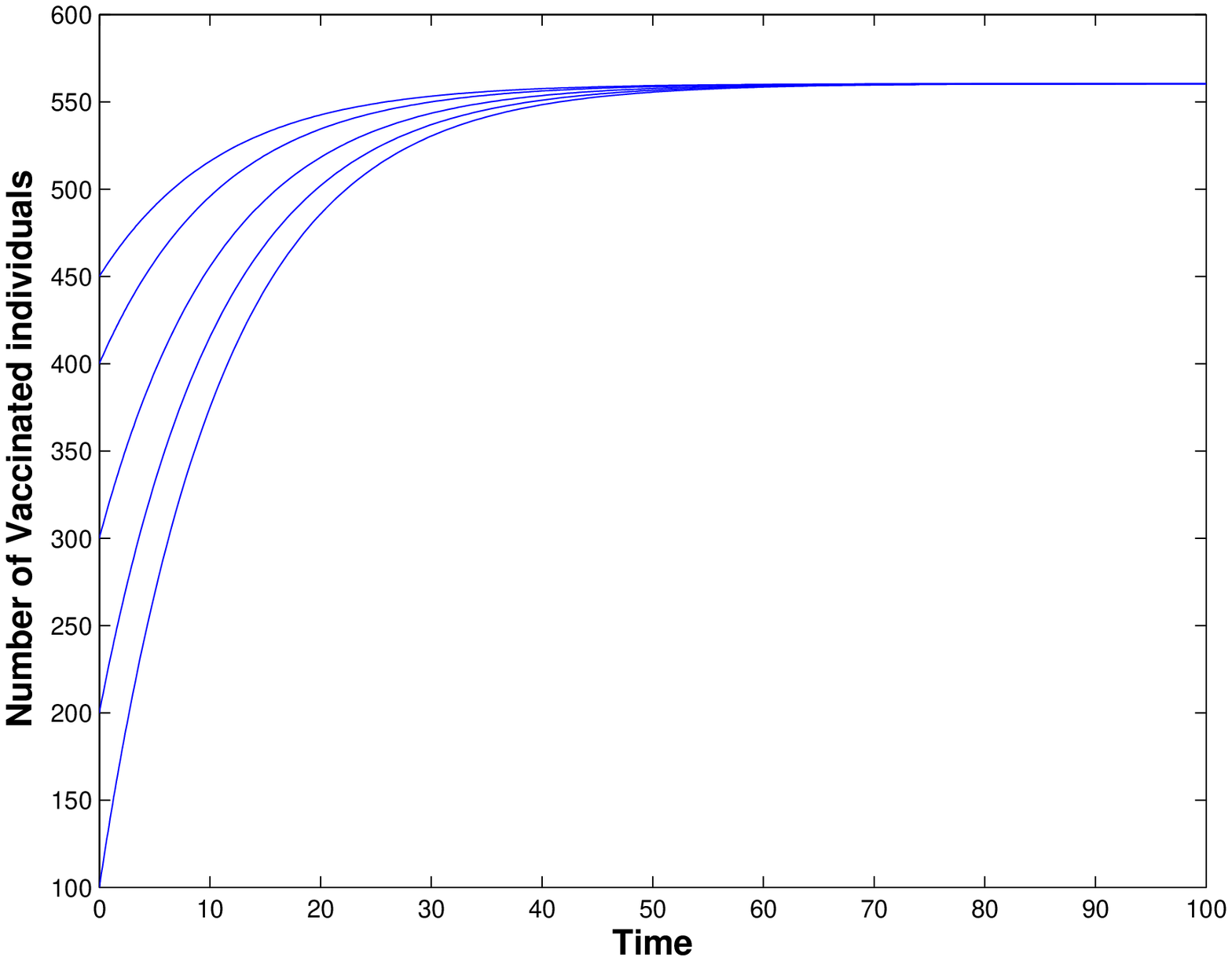}
\caption{Simulated in MATLAB, these figures from (a) to (f) are a simulation of system (\ref{eq3.8}), showing the total
number of susceptible, exposed, acutely infected, chronically
infected, treated and vaccinated individuals, respectively, as a function of time
(years). Parameter values are given in Table \ref{tab2}, with $\psi=1$ for
perfect vaccine, $\beta_1=0.0009,$ $\beta_2=0.0006,$
$\beta_3=0.0001$ and $\bar R_0=0.654<1.$ The numerical simulation
shows that the disease is eliminated when $\bar R_0<1$ .
 It is assumed that the acute
phase is more infectious than the chronic stage which is in turn
more infectious than the treatment phase. So
$\beta_1>\beta_2>\beta_3$.}
\label{a}
\end{figure*}

\vspace{.5em} \noindent \textbf{Proof}: Let

\[
V= A_1 E + A_2 I + A_3 T+ A_4 C_h,
\]
where
\[
A_1=\displaystyle\frac{S_0 \mu}{\Lambda},
A_2=\displaystyle\frac{S_0 K_1 \mu}{\epsilon \Lambda},
A_3=\displaystyle\frac{\beta_3S_0}{K_3},
A_4=\displaystyle\frac{\beta_2 S_0}{K_4}+\frac{\beta_3
S_0\pi_2}{K_3K_4}.
\]
Then,
\[
\begin{array}{rcl}
V^\prime&=&A_1E^\prime + A_2 I^\prime + A_3 T^\prime+ A_4
C_h^\prime\cr\cr &=&A_1\Big[(\beta_1 I+\beta_2 C_h+\beta_3 T)S-K_1
E\Big]+A_2\Big[\epsilon E-K_2 I\Big]+A_3\Big[\pi_1\kappa
I+\pi_2C_h-K_3T\Big]\cr\cr &&+A_4\Big[(1-\pi_1)\kappa I-K_4
C_h\Big].\end{array}
\]
Since, $S+E+I+T+C_h+V\leq\frac{\Lambda}{\mu},$ we have that
\[
S\leq\frac{\Lambda}{\mu}.
\]
Therefore $V^\prime$ becomes
\[
\begin{array}{rcl}
V^\prime&\leq &A_1\Bigr[(\beta_1 I+\beta_2 C_h+\beta_3
T)\displaystyle\frac{\Lambda}{\mu}-K_1 E\Bigr]+A_2\Big[\epsilon
E-K_2 I\Big]+A_3\Big[\pi_1\kappa I+\pi_2 C_h-K_3T\Big]\cr\cr
&&+A_4\Big[(1-\pi_1)\kappa I-K_4
C_h\Big]\cr\cr&=&E\Big[-K_1A_1+\epsilon A_2\Big]+I\Bigr[\beta_1 A_1
\displaystyle\frac{\Lambda}{\mu}-K_2A_2+\pi_1 \kappa A_3+A_4\kappa
(1-\pi_1)\Bigr]+T\Bigr[\beta_3A_1
\displaystyle\frac{\Lambda}{\mu}\cr\cr&&-K_3 A_3\Bigr] +
C_h\Bigr[\beta_2 A_1 \displaystyle\frac{\Lambda}{\mu}+\pi_2
A_3-K_4A_4\Bigr]\cr\cr&=& I \Bigr[S_0\Big(\beta_1 + \beta_2 \displaystyle\frac{\kappa (1 -
\pi_1)}{ K_4 } + \beta_3 \displaystyle\frac{( \pi_1 \kappa K_4 +
\pi_2 \kappa(1 -
\pi_1))}{K_3K_4}\Big)-\displaystyle\frac{S_0K_2K_1\mu}{\epsilon
\Lambda}\Bigr]\cr\cr&=&
\displaystyle\frac{IK_1K_2}{\epsilon}\Bigr[\bar R_0
-\displaystyle\frac{S_0\mu}{\Lambda}\Bigr]\leq 0,
\end{array}
\]
whenever
\[
\bar R_0\leq\displaystyle\frac{S_0\mu}{\Lambda}<1.
\]
Hence, $V^\prime\leq 0$ for $\bar
R_0\leq\displaystyle\frac{S_0\mu}{\Lambda}$. It should also be noted
that
$\displaystyle\frac{S_0\mu}{\Lambda}=\displaystyle\frac{\frac{\Lambda}{\mu}-\frac{b\Lambda}{\alpha+\mu}}{\frac
{\Lambda}{\mu} }<1.$  $V^\prime= 0$ whenever $E=0$, $I=0$, $T=0$,
$C_h=0$, which corresponds to the set $\{(E,I,T,C_h):E=I=T=C_h=0\}$.
In this set, system (\ref{eq3.8}) is given as
\begin{equation}
\label{eq3.11}
\begin{array}{rcl}
\displaystyle\frac{dS}{dt} &=&(1-b)\Lambda+\alpha V-\mu S,\cr\cr
\displaystyle\frac{dE}{dt}&=&
\displaystyle\frac{dI}{dt}\hspace{1.2mm} =\hspace{1.2mm}
\displaystyle\frac{dT}{dt}\hspace{1.2mm}=\hspace{1.2mm}
\displaystyle\frac{dC_h}{dt}\hspace{1.2mm}=\hspace{1.2mm}0,\cr\cr
\displaystyle\frac{dV}{dt} &=& b\Lambda-(\alpha+\mu)V.
\end{array}
\end{equation}
When $t\rightarrow\infty$, the solution of
system (\ref{eq3.11}) becomes
\[
S=\displaystyle \frac{(1-b)\Lambda}{\mu}+\frac{\alpha
b\Lambda}{\mu(\alpha+\mu)}, E=0, I=0, T=0, C_{h}=0, V=\frac{b\Lambda}{\alpha+\mu}.
\]
Clearly, when $t\rightarrow\infty,$ the solution to
system (\ref{eq3.11}) approaches the DFE, $P_0(S_0,0,0,0,0,V_0)$. By
using LaSalle's invariance principle (LaSalle 1976),  $P_0$ is found
to be globally asymptotically stable in $\Delta$.
This result is illustrated by simulating the model (\ref{eq3.8})
using a reasonable set of parameter values given in Table \ref{tab2}. The
plot in Fig. \ref{a} shows that the disease is eliminated from the population.

\subsection{Global stability of the endemic equilibrium}
\label{S7} 
\begin{theorem}
The endemic equilibrium $P^{\ast}(S^{\ast},E^{\ast},I^{\ast},$
$T^{\ast },C_{h}^{\ast },V^{\ast })$ of the system (\ref{eq2.1}),
with $\rho=0$ and $\sigma=0$, is globally asymptotically stable
whenever it exists.
\end{theorem}
In order to prove the above theorem, we have used the method given
by Li et al. (2012, 2011). At the endemic equilibrium $P^{\ast}$, with $\rho=0$
and $\sigma=0$, the following equations are satisfied:
\begin{equation}\label{eq4.10}
\begin{array}{ll} 
0=\displaystyle (1-b)\Lambda+\alpha
V^*-(\beta_{1}I^*+\beta _{2}C_{h}^*+\beta_{3}T^*)S^*-\mu S^*\medskip\\
\displaystyle 0=(\beta_{1}I^*+\beta_{2}C_{h}^*+\beta_{3}T^*)S^*+(1-\psi)
(\beta_{1}I^*+\beta_{2}C_{h}^*+\beta_{3}T^*)V^*-(\epsilon+\mu)E^*\medskip\\
\displaystyle 0=\epsilon E^*-(\kappa+\mu)I^*\medskip\\
 \displaystyle 0=\pi_{1}\kappa
I^*+\pi_{2}C_{h}^*-\mu T^*\medskip\\
 \displaystyle 0=(1-\pi_{1})\kappa
I^*-(\pi_{2}+\mu)C_{h}^*\medskip\\
\displaystyle 0=b\Lambda-(\alpha+\mu)V^*-(1-\psi)(\beta _{1}I^*+\beta
_{2}C_{h}^*+\beta_{3}T^*)V^*.
\end{array}
\end{equation}
Let
\begin{equation}\label{let}
x_1=\displaystyle\frac{S}{S^*},
\hspace{.2cm}x_2=\displaystyle\frac{E}{E^*},
\hspace{.2cm}x_3=\displaystyle\frac{I}{I^*},
\hspace{.2cm}x_4=\displaystyle\frac{T}{T^*},
\hspace{.2cm}x_5=\displaystyle\frac{C_h}{C_h^*},
\hspace{.2cm}x_6=\displaystyle\frac{V}{V^*}.
\end{equation}
Then (\ref{eq2.1}) can be rewritten as
\begin{equation}
\label{eq4.7}
\begin{array}{rcl}
x_1^\prime &=&
x_1\Bigr[\displaystyle\frac{(1-b)\Lambda}{S^*}\bigg(\frac{1}{x_1}-1\bigg)+\frac{\alpha
V^*}{S^*}\bigg(\frac{x_6}{x_1}-1\bigg)-\beta_{1}I^*(x_3-1)-\beta_{2}C_{h}^*(x_5-1)\cr\cr&&-\beta_{3}T^*(x_4-1)\Bigr],\cr\cr
x_2^\prime &=&
x_2\Bigr[\displaystyle\frac{\beta_{1}I^*S^*}{E^*}\bigg
(\frac{x_3x_1}{x_2}-1\bigg)
+\frac{\beta_{2}C_{h}^*S^*}{E^*}\bigg(\frac{x_1x_5}{x_2}-1\bigg)+
\frac{\beta_{3}T^*S^*}{E^*}\bigg(\frac{x_1x_4}{x_2}-1\bigg)\cr\cr
&&+ (1-\psi)\displaystyle\frac{\beta_{1}I^*V^*}{E^*}\bigg(
\frac{x_3x_6}{x_2}-1\bigg)+
(1-\psi)\frac{\beta_{2}C_{h}^*V^*}{E^*}\bigg(\frac{x_5x_6}{x_2}-1\bigg)+
(1-\psi)\cr\cr&&
\displaystyle\frac{\beta_{3}T^*V^*}{E^*}\bigg(\frac{x_4x_6}{x_2}-1\bigg)\Bigr],\cr\cr
x_3^\prime &=&
 x_3\Bigr[\displaystyle\frac{\epsilon
E^*}{I^*}\bigg(\frac{x_2}{x_3}-1\bigg)\Bigr],\cr\cr 
x_4^\prime &=&
x_4\Bigr[\displaystyle\frac{\pi_1\kappa
I^*}{T^*}\bigg(\frac{x_3}{x_4}-1\bigg)+\frac{\pi_2C_{h}^*}{T^*}\bigg(\frac{x_5}{x_4}-1\bigg)\Bigr],\cr\cr\\
x_5^\prime&=&x_5\Bigr[(1-\pi_1)\displaystyle\frac{\kappa
I^*}{C_h^*}\big(\frac{x_3}{x_5}-1\big)\Bigr],\cr\cr
x_6^\prime&=&
x_6\Bigr[\displaystyle\frac{b\Lambda}{V^*}\big(\frac{1}{x_6}-1\big)-(1-\psi)\beta_1I^*(x_3-1)-
(1-\psi)\beta_2C_h^*(x_5-1)-(1-\psi)\cr\cr&&
\beta_3T^*(x_4-1)\Bigr].
\end{array}
\end{equation}\normalsize
The endemic equilibrium $P^{\ast}(S^{\ast},E^{\ast},I^{\ast},$
$T^{\ast },C_{h}^{\ast },V^{\ast })$ corresponds to the positive
equilibrium $\bar P^*(1, 1, 1, 1,1,1)$ of (\ref{eq4.7}). Since, the
global stability of $\bar P^*$  is the same as that of $P^*$, the
global stability of $\bar P^*$ is described below instead of $P^*$.
We define the Lyapunov function as follows
\[
\begin{array}{rcl}
L &=& a_1 S^*(x_1-1-\mathrm{ln}x_1)+a_2 E^*(x_2-1-\mathrm{ln}x_2)+a_3
I^*(x_3-1-\mathrm{ln}x_3)\cr\cr && +a_4T^*(x_4-1-\mathrm{ln}x_4)+a_5
C_h^*(x_5-1-\mathrm{ln}x_5)+a_6 V^*(x_6-1-\mathrm{ln}x_6),
\end{array}
\]
where $a_1,a_2,a_3,a_4,a_5$ and $a_6$ are positive numbers which are
to be determined. Using (\ref{eq4.10}), the time derivative of $L$
along the solutions of system (\ref{eq2.1}) is given as
\begin{small}
\[
\begin{array}{rcl}
\displaystyle\frac{dL}{dt}&=&
 a_1\big(2(1-b)\Lambda  +\alpha V^* -\beta_1I^* S^*-\beta_2C_h^*S^*-\beta_3 T^*S^*\big)+a_2\big(\beta_1I^* S^*\cr\cr &&
+\beta_2C_h^*S^*+\beta_3T^*S^*+(1-\psi)\beta_1I^*V^*+(1-\psi)\beta_2C_h^*V^*+(1-\psi)\cr\cr &&
\beta_3T^*V^*\big)+a_3\epsilon E^*+a_4\pi_1\kappa I^*+ a_5(1-\pi_1)\kappa I^*+a_6\big(2b\Lambda-(1-\psi)\cr\cr &&
\beta_1I^*V^*-(1-\psi)\beta_2C_h^*V^*-(1-\psi)\beta_3T^*V^*\big)-x_1\big(a_1(1-b)\Lambda\cr\cr &&
+a_1\alpha V^*-a_1\beta_1I^*S^*-a_1\beta_2C_h^*S^*-a_1\beta_3T^*S^*\big)+x_2\big(-a_2\beta_1I^*S^*\cr\cr &&
-a_2\beta_2C_h^*S^*-a_2\beta_3T^*S^*-a_2(1-\psi)\beta_1I^*V^*-a_2(1-\psi)\beta_2C_h^*S^* \cr\cr &&
 -a_2(1-\psi)\beta_3T^*V^*+a_3\epsilon E^*\big)+x_3\big(a_1\beta_1I^*S^*-a_3\epsilon E^*+a_4\pi_1\kappa I^*\cr\cr &&
+a_5(1-\pi_1\kappa I^*+a_6(1-\psi)\beta_1I^*V^*)\big)+ x_4\big(a_1\beta_3T^*S^*-a_4\pi_1 \kappa I^* \cr\cr && 
-a_4\pi_2 C_h^*+a_6(1-\psi)\beta_3T^*V^*\big)+x_5\big(a_1\beta_2C_h^*S^*+a_4\pi_2C_h^*-a_5(1\cr\cr &&
-\pi_1)\kappa I^*+a_6(1-\psi)\beta_2C_h^*V^*\big)-x_6\big(-a_1\alpha V^*+a_6b\Lambda-a_6(1-\psi)\cr\cr &&
 \beta_1I^*V^*-a_6(1-\psi)\beta_2C_h^*V^*-a_6(1-\psi)\beta_3T^*V^*\big)+x_5x_6\big(a_2(1-\psi)\cr\cr &&
\beta_2C_h^*V^*-a_6(1-\psi)\beta_2C_h^*V^*\big)+x_1x_3\big(-a_1\beta_1I^*S^*+a_2\beta_1I^*S^*\big)\cr\cr &&
+x_1x_5\big(-a_1\beta_2 C_h^*S^*+a_2\beta_2C_h^*S^*\big)+x_1x_4\big(-a_1\beta_3T^*S^*+a_2\beta_3T^*S^*\big)\cr\cr&&+x_3x_6\big(a_2(1-\psi)\beta_1I^*V^*-a_6(1-\psi)\beta_1I^*V^*\big)+x_4x_6\big(a_2(1-\psi)\beta_3T^*V^*\cr\cr &&
-a_6(1-\psi)\beta_3T^*V^*\big)+\displaystyle\frac{1}{x_1}\big(-a_1(1-b)\Lambda\big)+\displaystyle\frac{1}{x_6}\big(-a_6b\Lambda\big)+\frac{x_6}{x_1}\big(-a_1\alpha \cr\cr &&
V^*\big)+\displaystyle\frac{x_3}{x_5}\big(-a_5(1-\pi_1\kappa I^*)\big)+\displaystyle\frac{x_5}{x_4}\big(-a_4\pi_2C_h^*\big)+\displaystyle\frac{x_3}{x_4}\big(-a_4\pi_1\kappa I^*\big)+\displaystyle\frac{x_2}{x_3}\cr\cr && 
\big(-a_3\epsilon E^*\big)+\displaystyle\frac{x_3x_6}{x_2}\big(-a_2(1-\pi)\beta_1 I^*V^*\big)+\displaystyle\frac{x_3x_1}{x_2}\big(-a_2\beta_1 I^*S^*\big)+\displaystyle\frac{x_5x_1}{x_2}\cr\cr &&
\big(-a_2\beta_2 C_h^*S^*\big)+\displaystyle\frac{x_1x_4}{x_2}\big(-a_2\beta_3 T^*S^*\big)+\displaystyle\frac{x_4x_6}{x_2}\big(-a_2(1-\pi)\beta_3 T^*V^*\big)\cr\cr &&
+\displaystyle\frac{x_5x_6}{x_2}\big(-a_2(1-\pi)\beta_2 C_h^*V^*\big)\cr\cr&=:& F(x_1,x_2,x_3,x_4,x_5,x_6).
\end{array}
\]
\end{small}
We define the function $H=\sum_{i=1}^{14} P_i$, where $P_i
(i=1,2,...,14)$ is given as

\begin{equation}\label{eq4.9}
\begin{array}{ll}
\displaystyle P_1=b_1\Big(2-x_1-\displaystyle\frac{1}{x_1}\Big),\vspace{.7em}\\
\displaystyle P_2=b_2\Big(2-x_6-\displaystyle\frac{1}{x_6}\Big),\vspace{.7em}\\
\displaystyle P_3=b_3\Big(3-x_1-\displaystyle\frac{1}{x_6}-\displaystyle\frac{x_6}{x_1}\Big),\vspace{.7em}\\
\displaystyle P_4=b_4\Big(3-\displaystyle\frac{1}{x_1}-\displaystyle\frac{x_1x_3}{x_2}-\displaystyle\frac{x_2}{x_3}\Big),\vspace{.5em}\\
\displaystyle P_5=b_5\Big(4-\displaystyle\frac{1}{x_1}-\displaystyle\frac{x_1x_4}{x_2}-\displaystyle\frac{x_3}{x_4}-
\displaystyle\frac{x_2}{x_3}\Big),\vspace{.7em}\\
\displaystyle P_6=b_6\Big(4-\displaystyle\frac{1}{x_1}-\displaystyle\frac{x_1x_5}{x_2}-\displaystyle\frac{x_3}{x_5}-
\displaystyle\frac{x_2}{x_3}\Big),\vspace{.7em}\\
\displaystyle P_7=b_7\Big(5-\displaystyle\frac{1}{x_1}-\displaystyle\frac{x_1x_4}{x_2}-\displaystyle\frac{x_5}{x_4}-
\displaystyle\frac{x_2}{x_3}-\frac{x_3}{x_5}\Big),\vspace{.7em}\\
\displaystyle P_8=b_8\Big(3-\displaystyle\frac{1}{x_6}-\displaystyle\frac{x_3x_6}{x_2}
-\displaystyle\frac{x_2}{x_3}\Big),\vspace{.5em}\\
\displaystyle P_9=b_9\Big(4-\displaystyle\frac{1}{x_6}-\displaystyle\frac{x_4x_6}{x_2}-\displaystyle\frac{x_3}{x_4}-
\displaystyle\frac{x_2}{x_3}\Big),\vspace{.7em}\\
\displaystyle P_{10}=b_{10}\Big(5-\displaystyle\frac{1}{x_6}-\displaystyle\frac{x_4x_6}{x_2}-\displaystyle\frac{x_5}{x_4}-
\displaystyle\frac{x_2}{x_3}-\frac{x_3}{x_5}\Big),\vspace{.7em}\\
\displaystyle P_{11}=b_{11}\Big(4-\displaystyle\frac{1}{x_6}-\displaystyle\frac{x_5x_6}{x_2}-\displaystyle\frac{x_3}{x_5}-
\displaystyle\frac{x_2}{x_3}\Big),\vspace{.7em}\\
\displaystyle P_{12}=b_{12}\Big(4-\displaystyle\frac{1}{x_6}-\displaystyle\frac{x_1x_3}{x_2}-\displaystyle\frac{x_6}{x_1}-
\displaystyle\frac{x_2}{x_3}\Big),\vspace{.7em}\\
\displaystyle P_{13}=b_{13}\Big(5-\displaystyle\frac{1}{x_6}-\displaystyle\frac{x_1x_5}{x_2}-\displaystyle\frac{x_6}{x_1}-
\displaystyle\frac{x_2}{x_3}-\frac{x_3}{x_5}\Big),\vspace{.7em}\\
\displaystyle P_{14}=b_{14}\Big(5-\displaystyle\frac{1}{x_6}-\displaystyle\frac{x_1x_4}{x_2}-\displaystyle\frac{x_6}{x_1}-
\displaystyle\frac{x_2}{x_3}-\frac{x_3}{x_4}\Big).\vspace{.7em}
\end{array}
\end{equation}
To determine all the coefficients, ( $a_i >0$ $(i=1,2,...,6)$, $b_i \geq 0$ $(i=1,2,...,14)$ ) we let
$F(x_1,x_2,x_3,x_4,x_5,x_6)$ = $H$. Comparing coefficients of $F$
and $H$, we see that the terms $x_2,~x_4,~x_5,~x_5x_6,~x_1x_3,~x_1x_5,~x_1x_4,~x_3x_6$, and
$x_4x_6$ of $F$ do not appear in $H$. Hence their coefficients will
be equal to zero. We solve the resulting equations to obtain
\[
a_1=a_2=a_6=1,
\]
\[ a_3=\displaystyle\frac{\epsilon+\mu}{\epsilon},
\]
\[
a_4=\displaystyle\frac{\beta_3S^*+(1-\psi)\beta_3V^*}{\mu},
\]
\[
a_5=\displaystyle\frac{\big(S^*+(1-\psi)V^*\big)\big(\beta_2+
\frac{\beta_3\pi_2}{\mu}\big)}{\pi_2+\mu}.
\]
Substituting these values into $L^\prime =
F(x_1,x_2,x_3,x_4,x_5,x_6)$, and using equations (\ref{eq4.10})
gives
\[
\begin{array}{rcl} &&
 F(x_1,x_2,x_3,x_4,x_5,x_6)=
\Big(2\Lambda+\alpha
V^*+(\epsilon+\mu)E^*+\beta_3\big(S^*+(1-\psi)V^*\big)T^*\cr\cr&&
+\big(S^*+(1-\psi)V^*\big)\big(\beta_2+ \displaystyle\frac{\beta_3
\pi_2}{\mu}\big)C_h^*\Big)- x_1\big(\mu S^*\big)-x_6\big(\mu
V^*\big)-\displaystyle\frac{1}{x_1} \big((1-b)\Lambda\big) \cr\cr &&-
\displaystyle\frac{1}{x_6}\big(b\Lambda\big)-\frac{x_6}{x_1}\big(\alpha
V^*\big)-\frac{x_3}{x_5}\Big(\big(S^*+(1-\psi)V^*\big)\big(\beta_2+
\displaystyle\frac{\beta_3
\pi_2}{\mu}\big)C_h^*\Big)-\displaystyle\frac{x_5}{x_4}
\Big(\displaystyle\frac{\beta_3}{\mu}\cr\cr &&\big(S^*+(1-\psi)V^*\big)\pi_2
C_h^*\Big)-
\displaystyle\frac{x_3}{x_4}\Big(\displaystyle\frac
{\beta_3}{\mu}\big(S^*+(1-\psi)V^*\big)\pi_1\kappa
I^*\Big)-\displaystyle\frac{x_2}{x_3}\Big((\epsilon+\mu)
E^*\Big)\cr\cr &&-\displaystyle\frac{x_3x_6}{x_2}\big((1-\psi)\beta_1
I^*V^*\big)-\displaystyle\frac{x_1x_3}{x_2}\big(\beta_1
I^*S^*\big)-\displaystyle\frac{x_1x_5}{x_2}\big(\beta_2
C_h^*S^*\big)-\displaystyle\frac{x_1x_4}{x_2}\big(\beta_3
T^*S^*\big)\cr\cr&&-\displaystyle\frac{x_4x_6}{x_2}\big((1-\psi)\beta_3
T^*V^*\big)-\displaystyle\frac{x_5x_6}{x_2}\big((1-\psi)\beta_2
C_h^*V^*\big).
\end{array}
\]
Comparing the remaining coefficients of $F$ and $H$ gives
\begin{equation}
\label{b11}
\begin{array}{ll}
b_1=\mu S^*-\alpha V^* +b_{12}+b_{13}+b_{14},\vspace{.5em}~~\\
b_2=\mu V^*\geq 0, \vspace{.5em}~~\\
b_3=\alpha V^* -b_{12}-b_{13}-b_{14},\vspace{.5em}~~\\
b_4=\beta_1 I^*S^* - b_{12},\vspace{.5em}~~\\
b_5=\beta_3 T^*S^*-\displaystyle\frac{\beta_3}{\mu}\big(S^*+(1-\psi)V^*\big)\pi_2C_h^*+b_{10}-b_{14},\vspace{.5em}~~\\
b_6=\beta_2C_h^*S^* -b_{13},\vspace{.5em}~~\\
b_7=\displaystyle\frac{\beta_3}{\mu}\big(S^*+(1-\psi)V^*\big)\pi_2C_h^* -b_{10},\vspace{.5em}~~\\
b_8=(1-\psi)\beta_1 I^* V^*\geq0, \vspace{.5em}~~\\
b_9=(1-\psi)\beta_3T^*V^* -b_{10}, \vspace{.5em}~~\\ 
b_{11}=(1-\psi)\beta_2 C_h^*V^*\geq0.
\vspace{.5em}
\end{array}
\end{equation}
To assure that $b_1, b_3, b_4, b_5, b_6, b_7$ and $b_9$ are non
negative, $b_{10},b_{12}, b_{13}, b_{14}$ must satisfy the following
inequalities 
\begin{equation}\label{eq101}
\begin{array}{ll}
\displaystyle \alpha V^*-\mu S^*\leq b_{12}+b_{13}+b_{14}\leq\alpha V^*,\vspace{.5em}\\
\displaystyle b_{10}\leq \min\Big((1-\psi)\beta_3 T^* V^*,
\displaystyle\frac{\beta_3}{\mu}\big(S^*+(1-\psi)V^*\big)\pi_2
C_h^*\Big),\vspace{.5em}\\
\displaystyle b_{14}-b_{10}\leq \beta_3 T^* S^*
-\displaystyle\frac{\beta_3}{\mu}\big(S^*+(1-\psi)V^*\big)\pi_2
C_h^*,\vspace{.5em}\\
\displaystyle b_{12}\leq \beta_1 I^*S^*,\vspace{.5em}
b_{13}\leq \beta_2 C_h^* S^*\vspace{.5em}.
\end{array}
\end{equation}\\
Finally, using equations (\ref{eq4.10}), the equality for the
constant terms between $F(x_1,x_2,x_3,x_4,x_5,x_6)$ and $H$ can easily be
verified.

\noindent The constrained conditions in (\ref{eq101}) show that the available
values of $b_{10}, b_{12}, b_{13},$ and $b_{14}$ are not unique.
Since, $b_1, b_3, b_4, b_5, b_6, b_7$ and $b_9$ depend on $b_{10},
b_{12}, b_{13},$ and $b_{14}$, their values will also be non unique.
Using inequalities in (\ref{eq101}), we can assign different values to
$b_i (i=1,3,...14, i\neq 2,8,11)$, and hence $H$ can have different
forms in following three subregions

\vspace{.2cm}

\noindent  \textbf{ Case 1}:  $\mu S > \alpha V,~~
\displaystyle\frac{\beta_3}{\mu}\big(S^*+(1-\psi)V^*\big)\pi_2
C_h^*\leq(1-\psi)\beta_3 T^* V^*.$

\vspace{.2cm}

\noindent For Case 1, using equations (\ref{b11}) and (\ref{eq101}), choose
$b_1=\mu S^*-\alpha V^*$, \vspace{.8em} $b_3=\alpha V^*$, 
$b_4=\beta_1 I^*S^*$, $b_5=\beta_3 T^* S^*$, $b_6=\beta_2 C_h^*
S^*$, $b_7=0$, $b_9=(1-\psi)\beta_3 T^* V^*\vspace{.2em} -
\displaystyle\frac{\beta_3}{\mu}\big(S^*+(1-\psi)V^*\big)\pi_2
C_h^*$,
$b_{10}=\displaystyle\frac{\beta_3}{\mu}\big(S^*+(1-\psi)V^*\big)\pi_2
C_h^*$,  $b_{12}=0$, $b_{13}=0$ \vspace{.4em} and  $b_{14}=0$.

\noindent Using these values, and the values of $b_2, b_8 $ and $b_{11}$, the
function $F(x_1,x_2,x_3,x_4,x_5,x_6)$ becomes

\vspace{.3cm}

$F(x_1,x_2,x_3,x_4,x_5,x_6)=$
\[
\begin{array}{ll}
\displaystyle (\mu S^*-\alpha V^*)\Big(2-x_1-\displaystyle\frac{1}{x_1}\Big)+\mu V^*\Big(2-x_6-\displaystyle\frac{1}{x_6}\Big)
\displaystyle +\alpha V^*\Big(3-x_1-\displaystyle\frac{1}{x_6} \vspace{.2cm}\\
-\displaystyle\frac{x_6}{x_1}\Big)\displaystyle +\beta_1 I^*
S^*\Big(3-\displaystyle\frac{1}{x_1}-\displaystyle\frac{x_2}{x_3}-\displaystyle
\frac{x_1x_3}{x_2}\Big) \displaystyle +\beta_3T^*S^*\Big(4-\displaystyle\frac{1}{x_1}-
\displaystyle\frac{x_2}{x_3}-\displaystyle \frac{x_1x_4}{x_2} \vspace{.2cm} \\ -
\displaystyle\frac{x_3}{x_4}\Big) \displaystyle +\beta_2 C_h^*S^*\Big(4-\displaystyle\frac{1}{x_1}-
\displaystyle\frac{x_2}{x_3}-\displaystyle \frac{x_1x_5}{x_2}- \displaystyle\frac{x_3}{x_5}\Big) \displaystyle +(1-\psi)\beta_1 I^*
V^*\Big(3-\displaystyle\frac{1}{x_6}\vspace{.2cm} \\
-\displaystyle\frac{x_2}{x_3}-\displaystyle
\frac{x_3x_6}{x_2}\Big)+\Big((1-\psi)\beta_3T^*
V^*\displaystyle -\displaystyle\frac{\beta_3}{\mu}\big(S^*+(1-\psi)V^*\big)\pi_2
C_h^*\Big)\Big(4-\displaystyle\frac{1}{x_6}\vspace{.2cm} \\-
\displaystyle\frac{x_2}{x_3}-\displaystyle \frac{x_4x_6}{x_2}-
\displaystyle\frac{x_3}{x_4}\Big) \displaystyle +\displaystyle\frac{\beta_3}{\mu}\big(S^*+(1-\psi)V^*\big)\pi_2
C_h^*\Big(5-\displaystyle\frac{1}{x_6}\displaystyle-
\displaystyle\frac{x_2}{x_3}-\displaystyle \frac{x_4x_6}{x_2}\vspace{.2cm} \\
-\displaystyle\frac{x_3}{x_5}
-\displaystyle\frac{x_5}{x_4}\Big) \displaystyle+(1-\psi
)\beta_2C_h^*V^*\Big(4-\displaystyle\frac{1}{x_6}-
\displaystyle\frac{x_2}{x_3}-\displaystyle \frac{x_5x_6}{x_2}-
\displaystyle\frac{x_3}{x_5}\Big).
\end{array}
\]

\noindent \textbf{Case 2}: $\mu S = \alpha V,~~
\displaystyle\frac{\beta_3}{\mu}\big(S^*+(1-\psi)V^*\big)\pi_2
C_h^*\geq(1-\psi)\beta_3 T^* V^*.$

\vspace{.2cm}

\noindent For Case 2, using equations (\ref{b11}) and (\ref{eq101}), choose
$b_1=0$,\vspace{.5em} $b_3=\alpha V^*$, $b_4=\beta_1 I^*S^*$,
$b_5=\beta_3\big(S^*+(1-\psi)V^*\big)\displaystyle\frac{\pi_1 \kappa
I^*}{\mu}$,  $b_6=\beta_2 C_h^* S^*$, \vspace{.5em}
$b_7=\displaystyle\frac{\beta_3}{\mu}\big(S^*+(1-\psi)V^*\big)\pi_2
C_h^*-(1-\psi)\beta_3 T^* V^*$, $b_9=0$, $b_{10}=(1-\psi)\beta_3
T^*V^*$, $b_{12}=0$,  $b_{13}=0$ and $b_{14}=0$.\\

\noindent Using the above values, and the values of $b_2, b_8 $ and
$b_{11}$, the function $F(x_1,x_2,x_3,x_4,x_5,x_6)$ becomes

\vspace{.3cm}

$F(x_1,x_2,x_3,x_4,x_5,x_6)$=
\[
\begin{array}{ll}
\mu V^*\Big(2-x_6-\displaystyle\frac{1}{x_6}\Big)+\alpha
V^*\Big(3-x_1-\displaystyle\frac{1}{x_6}-\displaystyle
\frac{x_6}{x_1}\Big) \displaystyle+\beta_1 I^*
S^*\Big(3-\displaystyle\frac{1}{x_1}-\displaystyle\frac{x_2}{x_3}-\displaystyle
\frac{x_1x_3}{x_2}\Big)\vspace{.2cm} \\
 \displaystyle+\beta_3\big(S^*+(1-\psi)V^*\big)\displaystyle\frac{\pi_1 \kappa
I^*}{\mu}\Big(4-\displaystyle\frac{1}{x_1}-
\displaystyle\frac{x_2}{x_3}-\displaystyle \frac{x_1x_4}{x_2}-
\displaystyle\frac{x_3}{x_4}\Big) \displaystyle+ \beta_2 C_h^*S^*\Big(4-\displaystyle\frac{1}{x_1}-
\displaystyle\frac{x_2}{x_3}\vspace{.2cm} \\
-\displaystyle \frac{x_1x_5}{x_2}-\displaystyle\frac{x_3}{x_5}\Big) \displaystyle+\big(\displaystyle\frac{\beta_3}{\mu}\big(S^*+(1-\psi)V^*\big)\pi_2 C_h^*-(1-\psi)\beta_3 T^* V^*\big) \displaystyle\Big(5-\displaystyle\frac{1}{x_1}-
\displaystyle\frac{x_2}{x_3}-\displaystyle \frac{x_1x_4}{x_2}\vspace{.2cm} \\-
\displaystyle\frac{x_3}{x_5}-\displaystyle\frac{x_5}{x_4}\Big) \displaystyle+(1-\psi)\beta_1 I^*
V^*\Big(3-\displaystyle\frac{1}{x_6}-\displaystyle\frac{x_2}{x_3}-\displaystyle
\frac{x_3x_6}{x_2}\Big) \displaystyle+(1-\psi)\beta_3T^*V^*\Big(5-\displaystyle\frac{1}{x_6}-
\displaystyle\frac{x_2}{x_3}\vspace{.2cm} \\
-\displaystyle \frac{x_4x_6}{x_2}-
\displaystyle\frac{x_3}{x_5}-\displaystyle\frac{x_5}{x_4}\Big) \displaystyle +(1-\psi
)\beta_2C_h^*V^*\Big(4-\displaystyle\frac{1}{x_6}-
\displaystyle\frac{x_2}{x_3}-\displaystyle \frac{x_5x_6}{x_2}-
\displaystyle\frac{x_3}{x_5}\Big).
\end{array}
\]

\begin{figure*}[!t]
\centering
\includegraphics[width=0.49\textwidth]{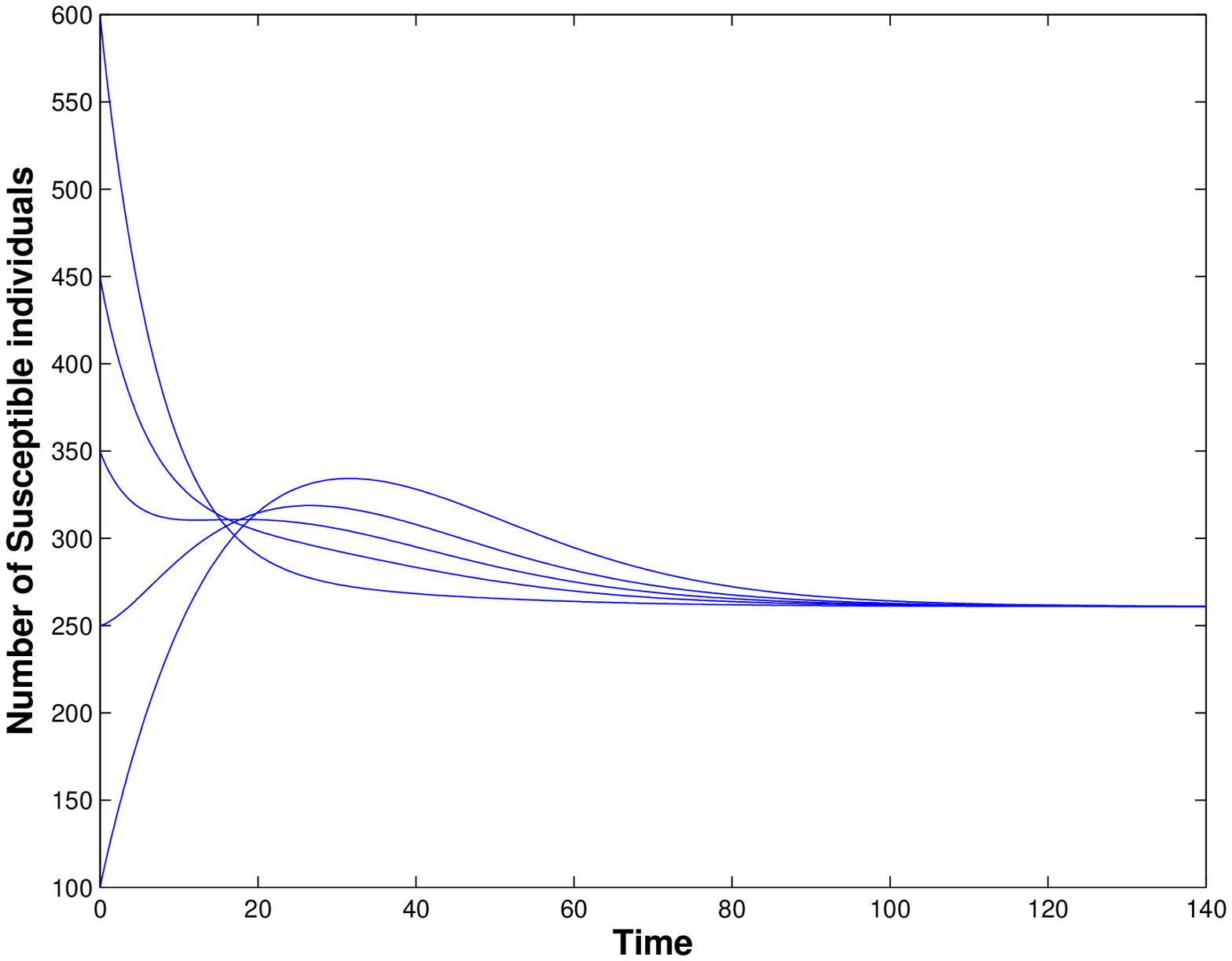}
\includegraphics[width=0.49\textwidth]{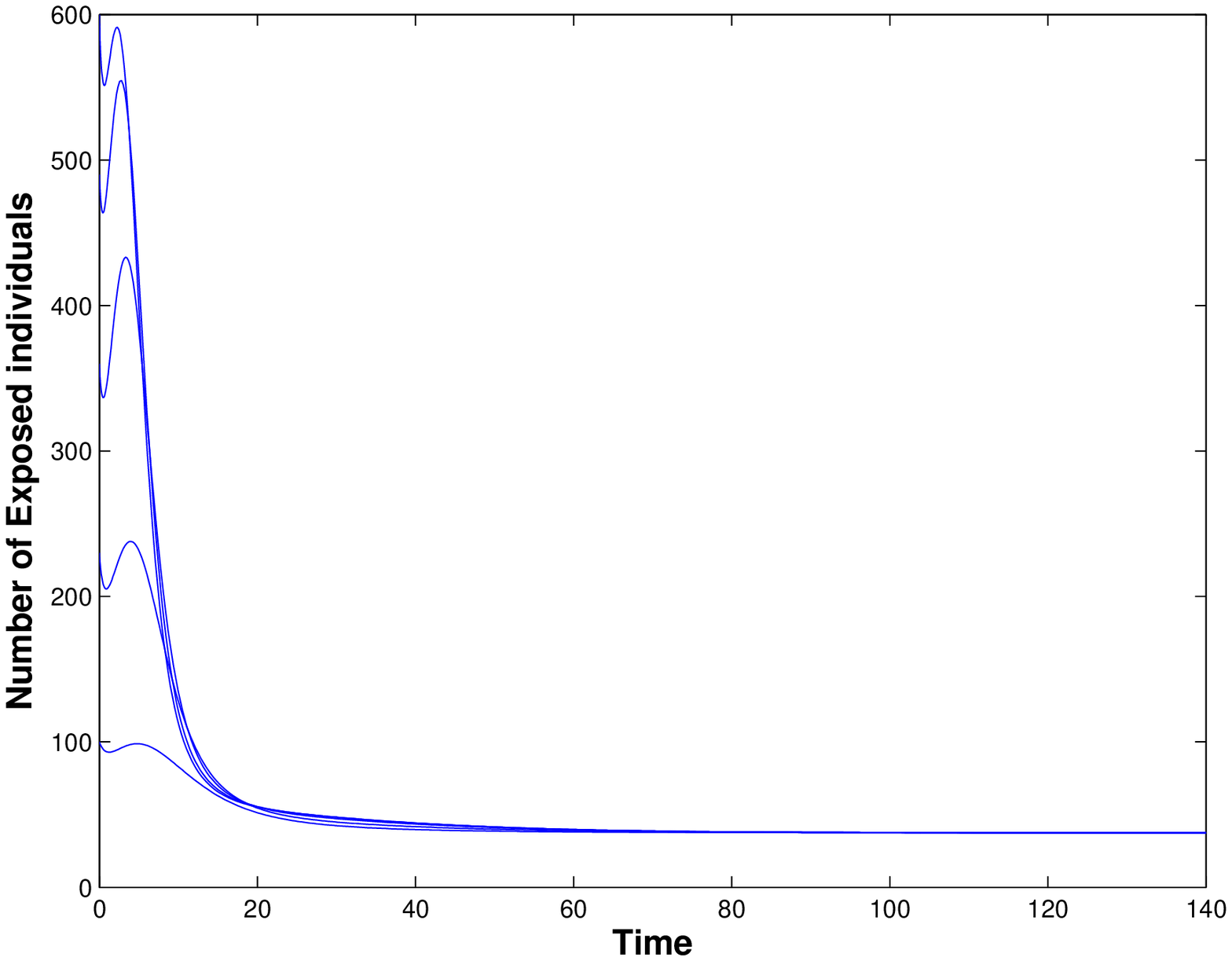}
\includegraphics[width=0.49\textwidth]{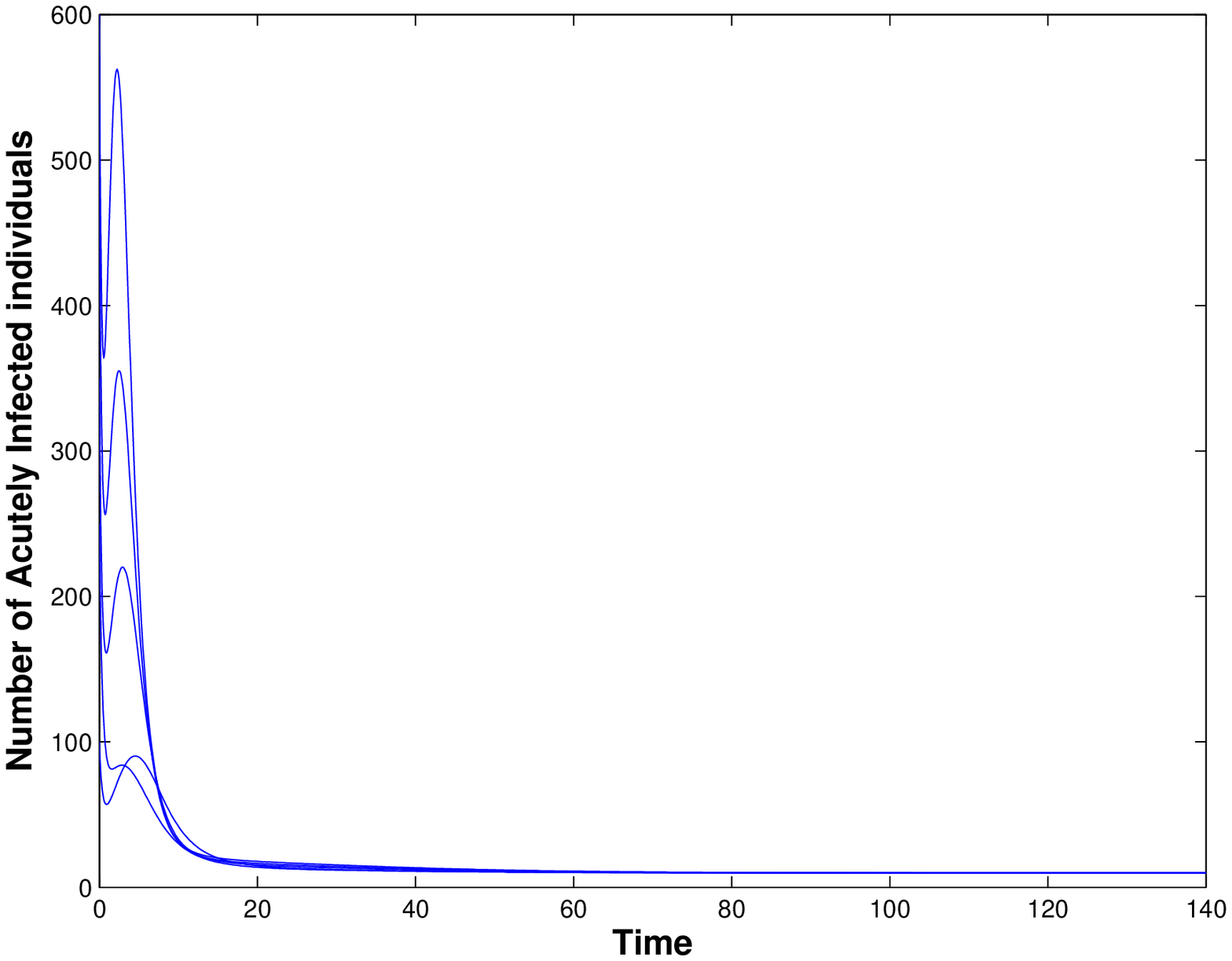}
\includegraphics[width=0.49\textwidth]{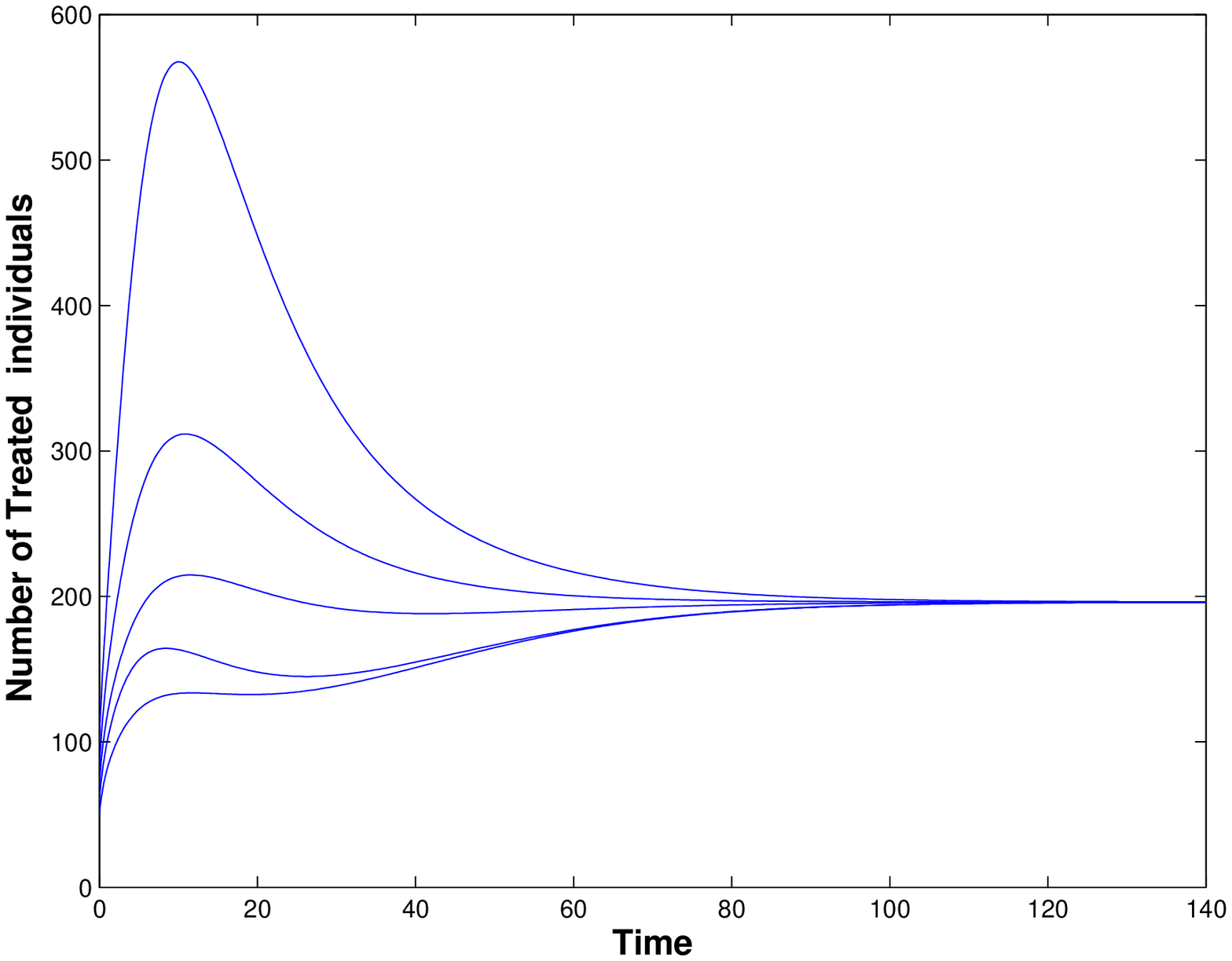}
\includegraphics[width=0.49\textwidth]{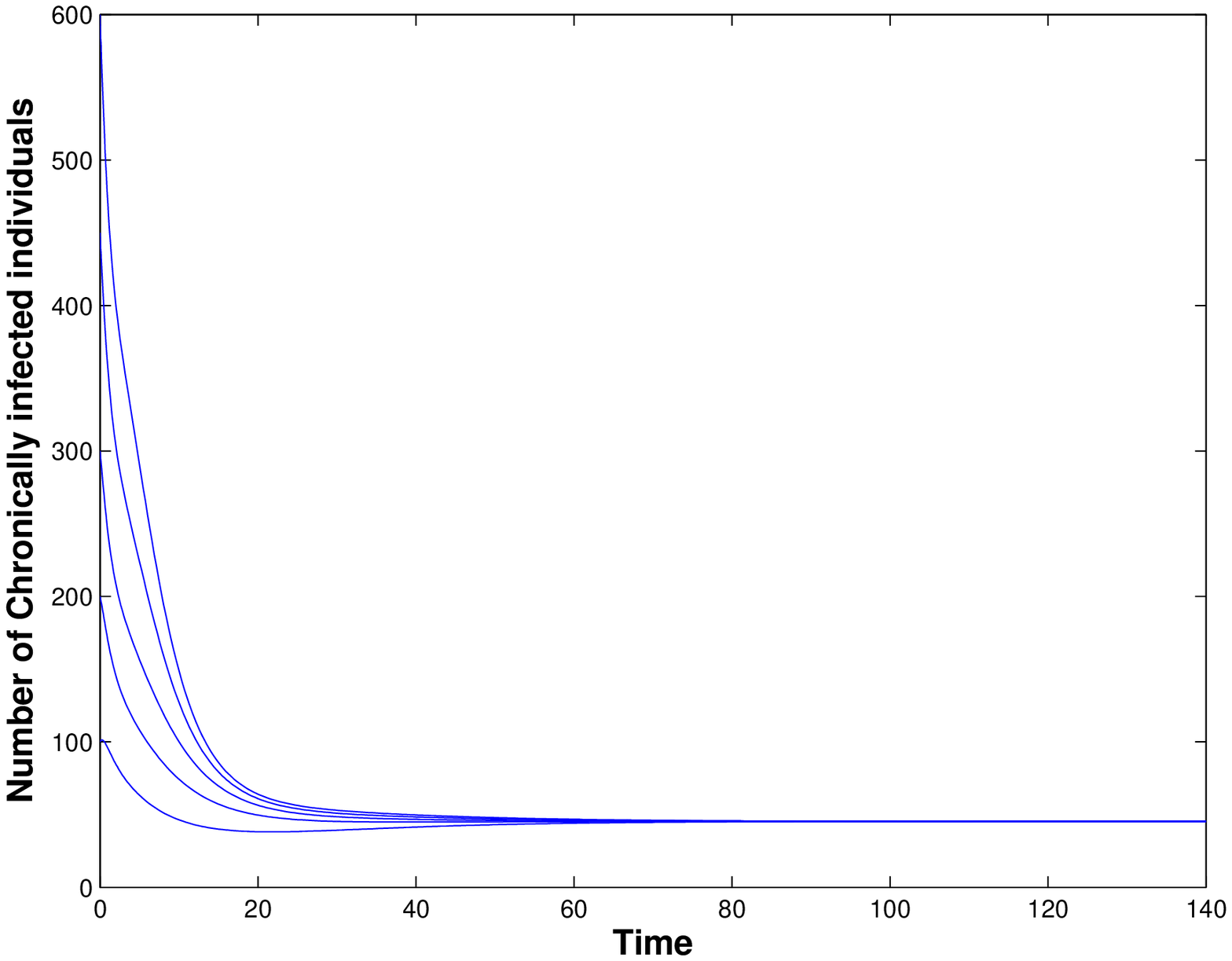}
\includegraphics[width=0.49\textwidth]{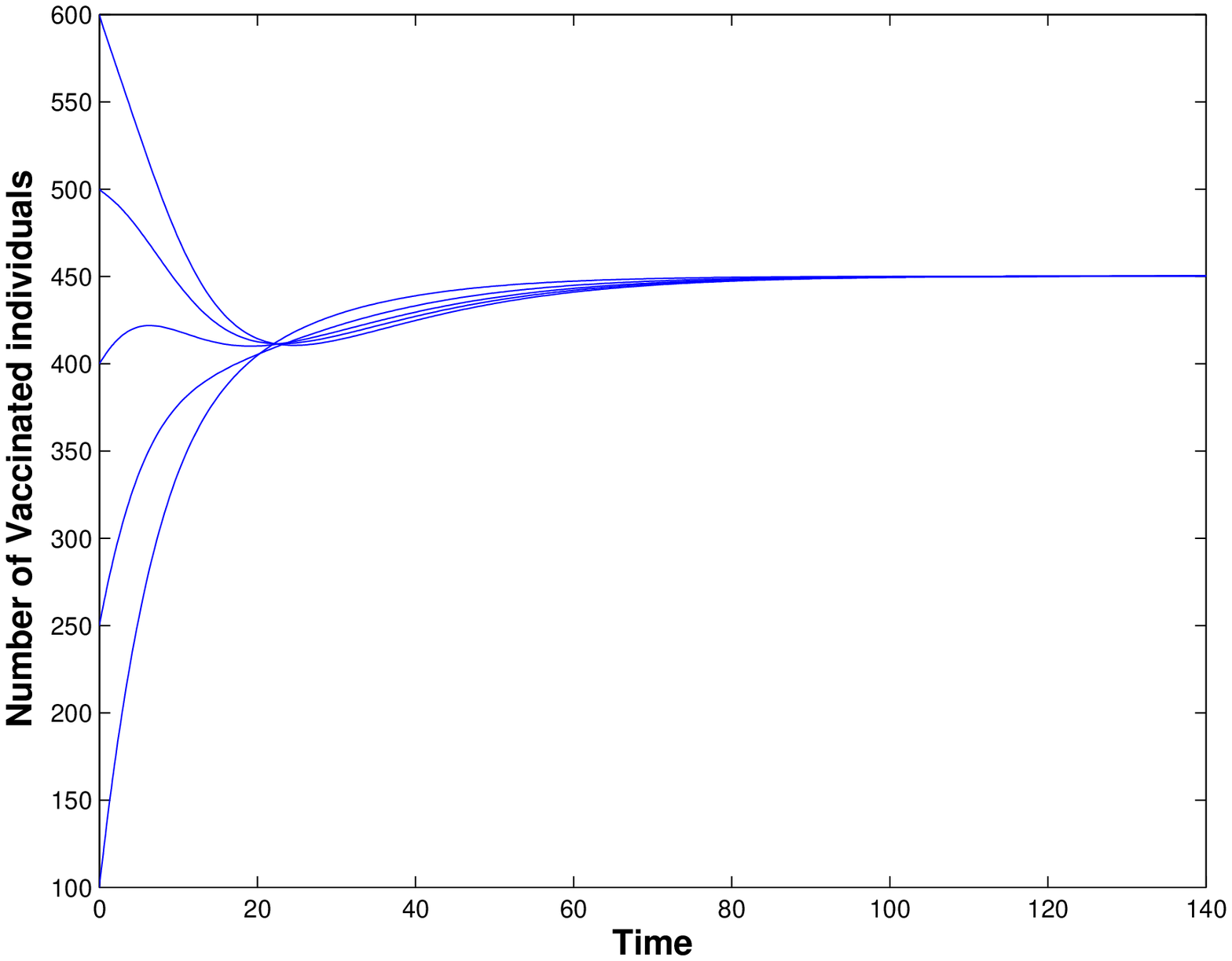}
\caption{Simulation (in MATLAB) of system (\ref{eq2.1}), showing the total
number of susceptible, exposed, acutely infected, chronically
infected, treated and vaccinated individuals as a function of time
(years) when $R_0>1$. Parameter values are given in Table \ref{tab2}, with
$\psi=0.6$, $\rho=0$, $\sigma=0$, $\beta_1=0.0009,$
$\beta_2=0.0006,$ and $\beta_3=0.0001$. The numerical simulation
shows that the disease persists when $R_0>1$.}
\label{b}
\end{figure*}

\noindent \textbf{Case 3}: $\mu S <\alpha V,~~
\displaystyle\frac{\beta_3}{\mu}\big(S^*+(1-\psi)V^*\big)\pi_2
C_h^*\geq(1-\psi)\beta_3 T^* V^*.$

\vspace{.2cm}

\noindent For Case 3, using equations (\ref{b11}) and (\ref{eq101}), we assume
that $\alpha
V^*\leq\beta_3\big(S^*+(1-\psi)V^*\big)\displaystyle\frac{\pi_1
\kappa I^*}{\mu}$ and choose $b_1=\mu S^*$,\vspace{.5em} $b_3=0$,
$b_4=\beta_1 I^*S^*$,
$b_5=\beta_3\big(S^*+(1-\psi)V^*\big)\displaystyle\frac{\pi_1 \kappa
I^*}{\mu}-\alpha V^*$,  $b_6=\beta_2 C_h^* S^*$, \vspace{.5em}
$b_7=\displaystyle\frac{\beta_3}{\mu}\big(S^*+(1-\psi)V^*\big)\pi_2
C_h^*-(1-\psi)\beta_3 T^* V^*$, $b_9=0$, $b_{10}=(1-\psi)\beta_3
T^*V^*$, $b_{12}=0$,  $b_{13}=0$ and $b_{14}=\alpha V^*$.\\

\noindent Using the above values, and the values of $b_2, b_8 $ and $b_{11}$,
the function $F(x_1,x_2,x_3,x_4,x_5,x_6)$ becomes\\

\vspace{.3cm}

$F(x_1,x_2,x_3,x_4,x_5,x_6)$=
\[
\begin{array}{lll}
\mu S^*\Big(2-x_1-\displaystyle\frac{1}{x_1}\Big)+\mu
V^*\Big(2-x_6-\displaystyle\frac{1}{x_6}\Big)+\beta_1 I^*
S^*\Big(3-\displaystyle\frac{1}{x_1}-\displaystyle\frac{x_2}{x_3}
-\displaystyle \frac{x_1x_3}{x_2}\Big)\vspace{.2cm} \\
+\Big(\beta_3\big(S^*+(1-\psi)V^*\big)\displaystyle\frac{\pi_1
\kappa I^*}{\mu}-\alpha V^*\Big)\Big(4-\displaystyle\frac{1}{x_1}-
\displaystyle\frac{x_2}{x_3}-\displaystyle \frac{x_1x_4}{x_2}-
\displaystyle\frac{x_3}{x_4}\Big)+\beta_2
C_h^*S^*\vspace{.2cm} \\
 \Big(4-\displaystyle\frac{1}{x_1}-
\displaystyle\frac{x_2}{x_3}-\displaystyle \frac{x_1x_5}{x_2}-
\displaystyle\frac{x_3}{x_5}\Big)+\displaystyle\frac{\beta_3}{\mu}\big(S^*+(1-\psi)V^*\big)\pi_2
C_h^*-(1-\psi)\beta_3 T^*V^* \vspace{.2cm} \\
\Big(5-\displaystyle\frac{1}{x_1}-
\displaystyle\frac{x_2}{x_3}-\displaystyle \frac{x_1x_4}{x_2}-
\displaystyle\frac{x_5}{x_4}-\displaystyle\frac{x_3}{x_5}\Big)
+(1-\psi)\beta_1 I^*
V^*\Big(3-\displaystyle\frac{1}{x_6}-\displaystyle\frac{x_2}{x_3}-\displaystyle
\frac{x_3x_6}{x_2}\Big) \vspace{.2cm} \\
+(1-\psi)\beta_3 T^*
V^*\Big(5-\displaystyle\frac{1}{x_6}-
\displaystyle\frac{x_2}{x_3}-\displaystyle \frac{x_4x_6}{x_2}-
\displaystyle\frac{x_3}{x_5}-\displaystyle\frac{x_5}{x_4}\Big)+(1-\psi
)\beta_2C_h^*V^*
\Big(4- \vspace{.2cm} \\ 
\displaystyle\frac{1}{x_6}-
\displaystyle\frac{x_2}{x_3}-\displaystyle \frac{x_5x_6}{x_2}-
\displaystyle\frac{x_3}{x_5}\Big)+\alpha
V^*\Big(5-\displaystyle\frac{1}{x_6}-
\displaystyle\frac{x_2}{x_3}-\displaystyle \frac{x_1x_4}{x_2}-
\displaystyle\frac{x_3}{x_4}- \displaystyle\frac{x_6}{x_1}\Big).
\end{array}
\]
Since, the arithmetic mean is greater than
or equal to the geometric mean, \\ $F(x_1,x_2,x_3,x_4,x_5,x_6)$ $\leq0$
in each of the above three cases. The equality holds only when
$x_1= x_6=1$, and $x_2=x_3=x_4=x_5$, i.e.
$\{(x_1,x_2,x_3,x_4,x_5,x_6)\in \Delta:
F(x_1,x_2,x_3,x_4,x_5,x_6)=0\}=
\{(x_1,x_2,x_3,x_4,x_5,x_6):x_1=x_6=1,x_2=x_3=x_4=x_5\}$.This
corresponds to the set $\Delta^\prime=\{(S,E,I,T,C_h,V):S=S^*,V=V^*,
E/E^*=I/I^*=T/T^*=C_h/C_h^*\}\in \Delta$. Hence, the maximum
invariant set of (\ref{eq2.1}) on the set $\Delta^\prime$ is the
singleton $\{P^*\}$. Therefore, by LaSalle's Invariance principle,
the endemic equilibrium $P^*$ is globally stable in $\Delta$ when
$\rho=0$ and $\sigma=0$. This result is illustrated by simulating the model in equation (\ref{eq2.1}) using a
reasonable set of parameter values given in Table \ref{tab2}. The plot shows
that the disease persists in the population (Fig. \ref{b}).

\section{Conclusions}
\label{S3} 
This paper presents a deterministic model for the transmission
dynamics of Hepatitis C virus infection.  The formulated model,
realistically, allows HCV transmission by acutely and chronically
infected individuals. Most importantly, the model includes a
compartment of vaccinated individuals, and considers the effect of a
waning vaccine on the transfer of individuals from one compartment
to another. The model was rigorously analyzed to gain insights into
its qualitative dynamics. We obtained the following results:
\begin{enumerate}
\item   The model has a locally stable disease free equilibrium whenever the associated
reproduction number is less than unity.
\item The model exhibits the
phenomenon of backward bifurcation, suggesting a case where stable
disease-free equilibrium co-exists with a stable endemic equilibrium
whenever the basic reproductive number is less than unity.
\item   Using an imperfect Hepatitis C vaccine would have no positive epidemiological impact to reduce disease burden in the community.
\item   Using a perfect vaccine can result in effective elimination of HCV infection in a
community, that is, the efficacy of the vaccine should be $100\%$
for complete removal of the disease.
\end{enumerate}%
\vspace{10pt} \noindent
%

%

\begin{thebibliography}{11}

\bibitem{Linda} Allen LJS (2007) An Introduction to Mathematical
Biology. Pearson Prentice Hall, New Jersey

\bibitem{Castillo} Castillo-Chavez C, Song B (2004) Dynamical model of tuberculosis and their
applications. Math Biosci Eng 1:361-404

\bibitem{Chen} Chen JY, Li F (2006) Development of hepatitis C virus vaccine using hepatitis B core antigen as
immuno-carrier. World J Gastroentero 12:7774-7778

\bibitem{Robin} Cotran RS, Kumar V, Robbins SL (1994) Pathologic basis of
disease. Saunders, Philadelphia

\bibitem{Dahari} Dahari H, Feliu A, Garcia-Retortillo M, Forns X, Neumann AU (2005) 
Second hepatitis C replication compartment indicated by viral dynamics during liver
transplantation. J Hepatol 42:491-498

\bibitem{Das}  Das P, Mukherjee D, Sarakar J (2005) Analysis of a disease
transmission model of hepatitis C. J Biol Syst 6:331-339

\bibitem{Bisceglie} Di Bisceglie AM (2000) Natural history of hepatitis C: its impact on
clinical management. Hepatology 31:1014-1018

\bibitem{Driessche} Driessche P, Watmough J (2002) Reproduction numbers and sub-threshold
 endemic equilibria for compartmental models of disease transmission.
Math Biosci 180:29-48

\bibitem{Elbasha} Elbasha EH (2013) Model for hepatitis C virus
 transmissions. Math Biosci Eng 10:1045-1065

\bibitem{4} Garba SM, Gumel AB, Abu Bakar MR (2008) Backward bifurcations in dengue transmission
dynamics. Math Biosci 215:11-25

\bibitem{Kretzschma} Jager J, Limburg W, Kretzschmar M, Postma M, Wiessing L (2004)
Hepatitis C and injecting drug use: impact, costs and policy options. Monograph, EMCDDA Lisbon

\bibitem{Lasalle} LaSalle JP (1976) The stability of dynamical systems.
Society for Industrial and Applied Mathematics, SIAM
Philadelphia 

\bibitem{9} Li J, Xiao Y, Zhang F, Yang Y (2012) An algebraic approach to proving the global
stability of a class of epidemic models. Nonlinear Anal-Real 13:2006-2016

\bibitem{Li} Li J, Yang Y, Zhou Y (2011) Global stability of an epidemic model with latent stage and
vaccination. Nonlinear Anal-Real 12:2163-2173

\bibitem{World} Lozano R, Naghavi M, Foreman K, Lim S, Shibuya K, Aboyans V, Abraham J (1990)
Hepatitis C factsheet no.164. World Health Organization. 
http://www.who.int/mediacentre/factsheets/fs164/en/. Accessed 10
December 2012

\bibitem{Martcheva} Martcheva M, Castillo-Chavez C (2003) Diseases with chronic stage in a population with varying
size. Math Biosci 182:1-25

\bibitem{num1} Martin NK, Vickerman P, Foster GR, Hutchinson SJ, Goldberg DJ, Hickman M (2011) Can antiviral therapy for
hepatitis C reduce the prevalence of HCV among injecting drug user
populations? A modeling analysis of its prevention utility. J Hepatol  54:1137-1144

\bibitem{num2} Martin NK, Vickerman P, Hickman M (2011) Mathematical modelling of hepatitis C treatment for injecting drug
users. J Theor Biol 274:58-66

\bibitem{Wu} Qesmi R, Wu J, Heffernan JM (2010) Influence of backward bifurcation
in a model of hepatitis B and C viruses. Math Biosci 224:118-125

\bibitem{Safi} Safi MA, Gumel AB (2011) Mathematical analysis of a disease transmission model with quarantine,
isolation and an imperfect vaccine. Comput Math Appl 61:3044-3070

\bibitem{Zeiler} Zeiler I, Langlands T, Murray JM, Ritter A (2010) Optimal
targeting of Hepatitis C virus treatment among injecting drug users
to those not enrolled in methadone maintenance programs. Drug
Alcohol Depen 110:228-233

\bibitem{5}  Zhang S, Zhou Y (2012) The analysis and application of an HBV
model. Appl Math Model 36:1302-1312



\end{thebibliography}
\end{document}